\documentclass[preprint,prd,tightenlines,nofootinbib]{revtex4}

\pdfoutput=1 
\usepackage{amsmath,bm}
\usepackage{graphicx}
\usepackage{hyperref}
\usepackage[usenames,dvipsnames]{color}

\newcommand{\bra}[1]{\langle #1|}
\newcommand{\ket}[1]{|#1\rangle}

\newcommand{\as}{\alpha_{\mathrm{s}}}

\newcommand{\LE}{\mathrm{E}}

\newcommand{\LI}{\mathrm{I}}

\newcommand{\LL}{\mathrm{L}}

\newcommand{\LT}{\mathrm{T}}

\newcommand{\Ld}{\mathrm{d}}

\newcommand{\Lq}{\mathrm{q}}
\newcommand{\Ls}{\mathrm{s}}
\newcommand{\Lu}{\mathrm{u}}

\newcommand{\GeV}{\ \mathrm{GeV}}
\newcommand{\MeV}{\ \mathrm{MeV}}
\begin{document}

\title{Scattering of Dark Particles with Light Mediators}

\author{Davison E.\ Soper}
\affiliation{
Institute of Theoretical Science\\
University of Oregon\\
Eugene, OR  97403-5203, USA\\
}
\author{Michael Spannowsky and Chris J.\ Wallace}
\affiliation{
Institute for Particle Physics Phenomenology\\
Department of Physics\\
Durham University\\
Durham CH1 3LE,
United Kingdom\\
}

\author{Tim M. P. Tait}
\affiliation{
Department of Physics\\
University of California, Irvine\\
Irvine, CA 92697, USA
}

\begin{abstract}
We present a treatment of the high energy scattering of dark Dirac fermions from nuclei, mediated by the exchange of a light vector boson.
The dark fermions are produced by proton-nucleus interactions in a fixed target and, after traversing shielding that screens out strongly interacting products, appear
similarly to neutrino neutral current scattering in a detector.
Using the Fermilab experiment E613 as an example, we place limits on a secluded dark matter scenario.
Visible scattering in the detector includes both the familiar regime of large momentum transfer to the nucleus ($Q^2$)
described by deeply inelastic scattering, as well as small $Q^2$ kinematics described by the exchanged vector mediator fluctuating into a
quark-antiquark pair whose interaction with the nucleus is described by a saturation model.
We find that the improved description of the low $Q^2$ scattering leads to important corrections, resulting in more robust
constraints in a regime where a description entirely in terms of deeply inelastic scattering cannot be trusted.
\end{abstract}

\preprint{UCI-HEP-TR-2014-04, IPPP/14/61, DCPT/14/122}

\maketitle

%-------------------------------------------------------------------

\section{Introduction and motivation}

There is compelling evidence that most of the mass in the Universe is in the form
of nonbaryonic dark particles.  And yet, the identity of this dark matter (DM) remains
elusive.  Among the many proposed candidates, weakly-interacting massive
particles (WIMPs) are the most popular, due to the fact that their abundance in the Universe can be explained by virtue of their being thermal relics provided they have weak scale masses and couplings~\cite{Bertone:2004pz}.

One possibility is that the dark matter particles do not interact with ordinary matter strictly by the weak force. Rather, they may be able to exchange particles that interact with quarks or gluons. In this case, the relevant couplings would have to be small. Such particles could potentially be discovered by any of three methods. First, dark matter particles in locations in our galaxy where they are especially abundant could annihilate to form baryonic matter and, eventually, photons that might be detected (indirect detection). Second, dark matter particles in the halo of our galaxy might interact with nuclei in a detector on earth and this interaction might be observable (direct detection). Third, dark matter particles might be created in hadron collisons at an accelerator (accelerator production). If this happens often enough at a colliding beam accelerator such as the Large Hadron Collider, one might discover these events by looking, for example, for a missing energy signal. Alternatively, one might create dark matter particles in hadron collisions with nuclei in a fixed target and detect them through their interactions with nuclei in a suitable detector.

Currently, the best constraints on dark particles interacting with quarks come from a mixture of searches for direct
detection and accelerator production.
In a direct detection experiment, a particle $\chi$ with mass $m_\chi$ and velocity $v_\chi$ interacts with a nucleus in the detector and one looks for the nuclear recoil, where the 
typical magnitude of $v_\chi \simeq 10^{-3}$ is determined by the gravitational potential of the Galaxy. If $m_\chi$ is not large enough, the momentum $m_\chi v_\chi$ will not be large enough to create an observable nuclear recoil \cite{Akerib:2013tjd,Aprile:2012nq,Ahmed:2009rh,Angloher:2011uu}. For this reason, the current generation of direct detection experiments have not been sensitive to dark matter particles with $m_\chi \lesssim 5$~GeV. However, these limits may improve in experiments using specialized detection techniques (e.g. based on measurements of ionization yield) \cite{Essig:2012yx,Cushman:2013zza}.
As a result, the best bounds on hadronic interactions for such light dark matter particles currently come from accelerator production at colliders \cite{Goodman:2010yf,Bai:2010hh,Goodman:2010ku,Fox:2011pm,Rajaraman:2011wf,Bai:2012xg,Aaltonen:2012jb,Cheung:2012gi,Chatrchyan:2012tea,Chatrchyan:2012pa,ATLAS:2012ky}, particularly for the case in which the particles mediating these interactions are heavy compared to the momentum transfer of the production process.

Of special interest are models in which the dark sector particles that mediate the interactions 
between the $\chi$ and standard model particles are not heavy but rather light, in some cases even
lighter than the $\chi$ particles. This is the secluded scenario of Refs.~\cite{Batell:2009yf,Batell:2009di}.  If the dark matter particles $\chi$ are themselves light enough so that they escape from direct detection experiments, a promising way to look for them is at fixed target experiments \cite{Batell:2009di,Essig:2010gu} 
where a beam of protons strike a target to produce a beam of $\chi$ particles which are sufficiently weakly interacting so as to pass through shielding (as do neutrinos)
where they can eventually be detected via their rare scattering with the nuclei comprising a detector.
The advantage of a fixed target experiment over a colliding beam experiment is the higher luminosity that a fixed target experiment can offer, a key factor when searching
for extremely rare production processes. 
In particular, we focus on the Fermilab beam dump experiment E613, which utilized a 400 GeV incoming proton beam on a tungsten target.
Future high energy beam dump experiments could potentially extend the reach of E613 \cite{Hewett:2014qja}.

We employ a very simple model for the dark sector of the theory consisting of
a single Dirac fermion dark matter particle $\chi$ and a light vector particle $V$, which couples to both $\chi$ particles as well as quarks. 
We refer to $V$ as the dark vector boson. The relevant interactions are
\begin{equation}
\mathcal{L}_\LI = V_\mu \left( g_{q\bar q v} ~ \sum_q ~ \bar{q} \gamma^\mu q
+ g_{\chi \bar \chi v} ~ \bar{\chi} \gamma^\mu \chi \right)~.
\label{eq:lag}
\end{equation}
This framework is similar to a ``dark photon" model, in which $V$ picks up interactions to the Standard Model through kinetic mixing with hyper-charge \cite{Dienes:1996zr}, but differs in that it has universal charges for the quarks and is agnostic concerning the coupling to leptons. We discuss the
dark photon case in more detail below, but it is worth noting here that for the regions of parameter space of interest to us, $1 \MeV < m_\chi < 10 \GeV$ and $m_v \sim 1 \MeV$, there
are much stronger constraints on a dark photon mediator from experiments with
electrons on fixed targets \cite{Bjorken:2009mm,Batell:2014mga} 
than on models interacting only with
quarks \cite{Carone:1994aa}. 
Thus one might consider the interaction (\ref{eq:lag}) in a leptophobic model in which the light vector particles do not couple to leptons.
The leptophobic model is not really intended to be taken as a realistic model
for the dark sector, but is a convenient framework to explore the degree to which non-perturbative
QCD plays a role in describing how $\chi$ particles scatter off of the nuclei in a detector. The high energy of the $\chi$ particles produced by E613's $400 \GeV$ beam demands this more detailed treatment of scattering than is necessary for the low energy neutrino factories discussed in the context of a similar model in \cite{deNiverville:2011it,deNiverville:2012ij,Batell:2014yra}.

We will frame the discussion in terms of a dark matter search at E613 using the simple model of Eq.~(\ref{eq:lag}). In Section \ref{sec:model} we describe the production of dark particles at proton fixed target experiments. 
In Section \ref{sec:scattering}, we calculate the rescattering rate of produced $\chi$s in the detector, using both a deeply inelastic scattering (DIS) approach, detailed in Section \ref{sec:DIS}, and a parton saturation approach, detailed in Section \ref{sec:SAT}. We examine the connection between the two approaches in Section \ref{sec:connection}. In Section \ref{sec:dm}, we use the results of experiment E613 to place limits on the couplings in Eq.~(\ref{eq:lag}) and in a closely related ``minicharge'' model.
Finally, we present conclusions in Section~\ref{sec:conclusions}.
Details of the kinematics are provided in an Appendix.

\section{Production of dark matter particles}
\label{sec:model}

\begin{figure}
\centering
\includegraphics[width=5 cm]{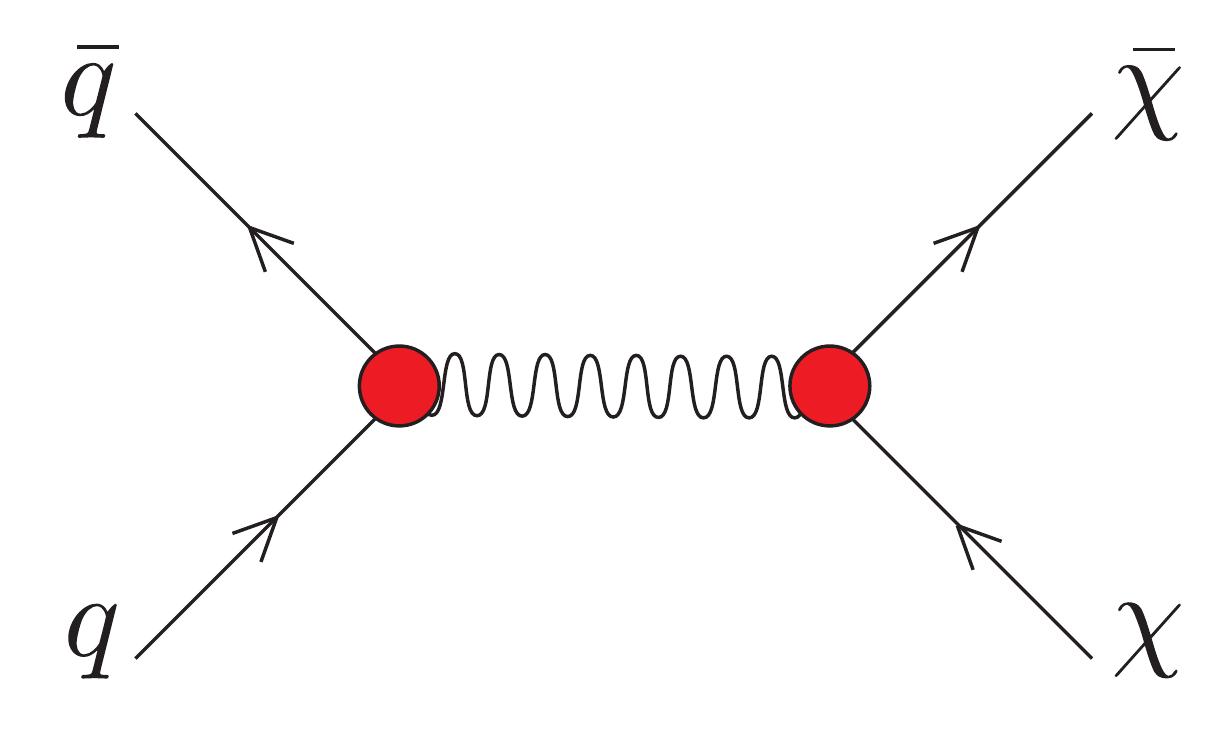}\\
\caption{Feynman diagram for direct production of $\chi$ particles from $p\,A$ collisions.}
\label{fig:chiproduction}
\end{figure}

When beam protons strike the tungsten target in experiment E613, they can produce $\chi \bar \chi$ pairs through the diagram shown in Fig.~\ref{fig:chiproduction}. We demand that one or both of the $\chi$ particles have a high energy in the lab frame. Then this is a hard process that can be reliably calculated in lowest order perturbation theory, taking the tungsten nucleus to consist of $Z = 74$ protons and $A - Z \approx 110$ neutrons, treated as non-interacting. The interactions of Eq.~(\ref{eq:lag}) are implemented in Madgraph 5 \cite{Alwall:2011uj} with the help of FeynRules \cite{Christensen:2008py}.  The inclusive cross sections for the process
\begin{equation}
p  p \rightarrow \bar{\chi} \chi + X
\end{equation}
for a proton of energy $E_B$ incident upon a proton at rest is simulated
at the parton level in the Monte Carlo generator.  In order to convert this into the number of $\chi$s or $\bar\chi$s produced with energy $E$ and angle $\theta$, we write (approximating the cross section from neutrons in the nucleus as being identical to the cross sections from protons, as is approximately true in our model (\ref{eq:lag}))
\begin{equation}
\frac{dN}{dE\,d\theta}  =  
A\, \frac{d\sigma (p p \rightarrow \chi \bar{\chi})}{d E\, d\theta}\,
L_T\, n_T \, {\rm POT}
~,
\end{equation}
multiplying by the length of the target $L_T$, the density of tungsten nuclei inside it, $n_T$, and the number of protons incident on the target corresponding to the data set, POT.
Here the cross section is the cross section to produce either a $\chi$ or a $\bar \chi$. 

\begin{figure}
\centering
\includegraphics[width=12 cm]{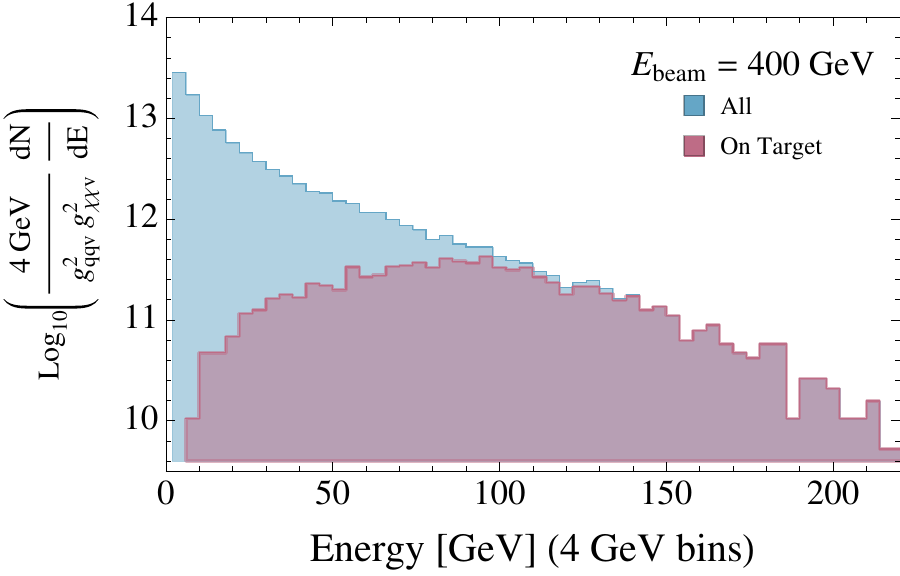}\\
\caption{Typical distribution of $\chi$ particles as a function of energy, $\frac{dN}{dE}$, divided by $g_{q \bar q v}^2\, g_{\chi \bar \chi v}^2$. The vertical scale is logarithmic. We show the distribution of all produced particles $\chi$ and $\bar \chi$ and the distribution of particles produced at angles that will result in their impacting the target. Many of the lowest energy dark particles are produced at wide angles and miss the detector.}
\label{fig:dNdE}
\end{figure}

The number of $\chi$s that actually make it to a detector further depends
on the angular acceptance of the detector.  The E613 detector geometry is somewhat complicated in this regard.  The detector face was $3~{\rm m} \times 1.5~{\rm m}$, with the beam offset along the horizontal axis by $0.75$~m.  To be conservative, we assume $\chi$s must be incident within the $0.75$~m radius circle centered on the beam axis, though in practice there
was a larger instrumented region which could be capable of detecting additional
$\chi$s with larger production angles.  The produced $\chi$s are thus incident on the detector provided their production angle is less than,
\begin{equation}
\theta_{\rm max}  =  \frac{0.75~{\rm m}}{55.8~{\rm m}} =  0.0134~.
\end{equation}
The number of $\chi$s per unit energy incident on the detector is then\footnote{Some dark matter particles can be lost on their way to the detector because they scatter in the rock that lies between the production point and the detector or in the iron shielding of the detector. We 
discuss this effect in the calculations of Sec.~\ref{sec:dm}.}
\begin{equation}
\label{eq:dNdE}
\frac{dN}{dE}  =  \int_0^{\theta_{\rm max}}\!d\theta\
\frac{dN}{dE\,d\theta}
~.
\end{equation}
In Fig.~\ref{fig:dNdE}, we show a plot of the calculated ${dN}/{dE}$ divided by $g_{q \bar q v}^2\, g_{\chi \bar \chi v}^2$.

\section{Structure functions for dark matter scattering in the detector}
\label{sec:scattering}

The detector is made of lead plus liquid scintillator. When a $\chi$ particle enters the detector with energy $E$, it can scatter from a lead nucleus. In order for the scattering to be detected, we demand that the scattering transfer at least an amount of energy $E_{\rm cut}$ to the nucleus. We take $E_{\rm cut} = 20 \GeV$, corresponding to the minimum energy demanded by the detector to register a jet \cite{Romanowski:1985xn,Duffy:1988rw}. Thus the expected number of events is proportional to the convolution of $dN/dE$ from Eq.~(\ref{eq:dNdE}) with the cross section $\sigma(E,E_{\rm cut})$ for a $\chi$ particle to deposit energy greater than $E_{\rm cut}$ in the nucleus.

\begin{figure}
\centering
\includegraphics[width = 8 cm]{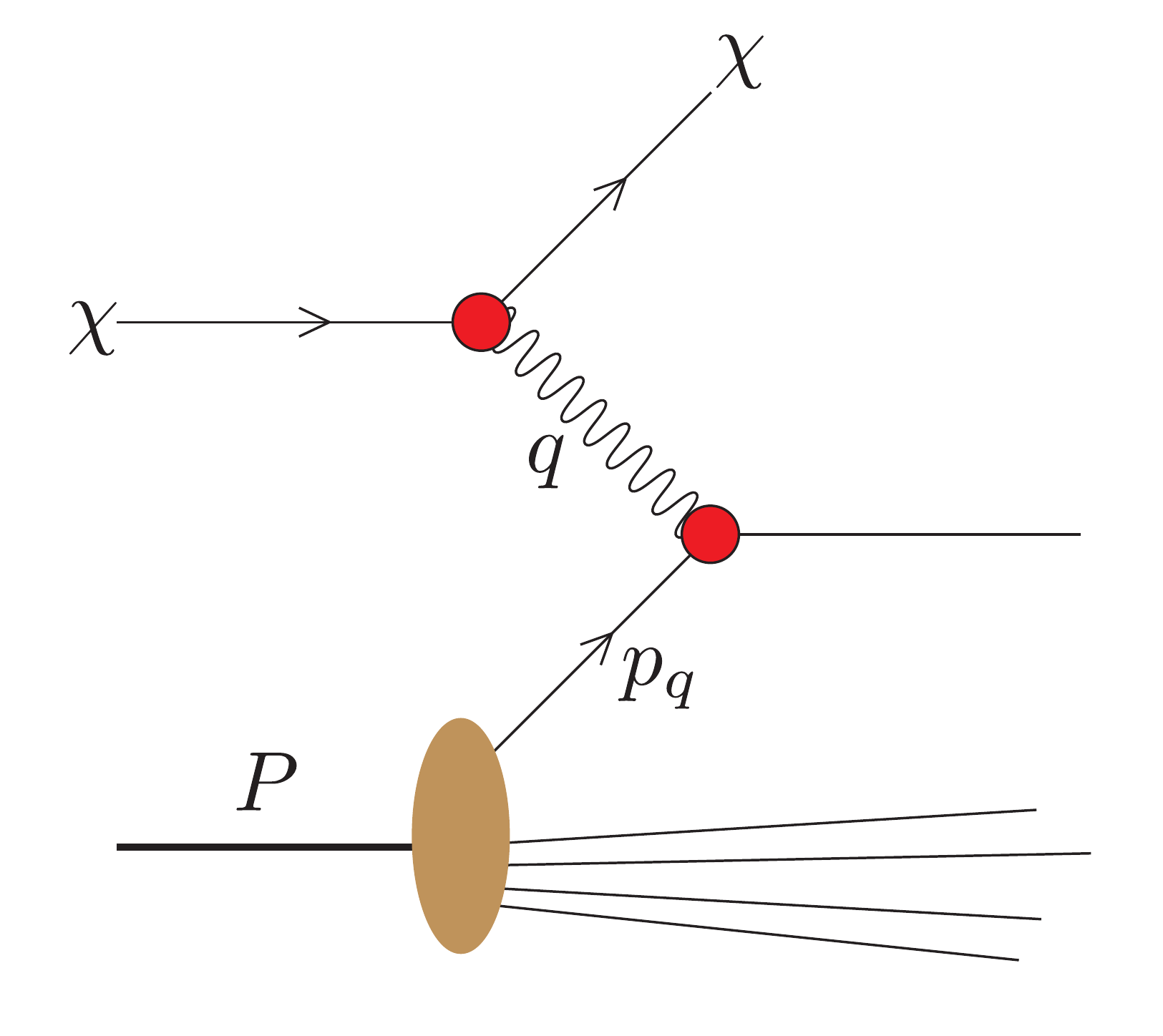}
\caption{Classic picture of deeply inelastic scattering from a lead nucleus, with the exchanged vector boson replaced by a massive dark vector boson that carries momentum $q$ and interacts with a quark from the nucleus carrying momentum $p_q$.}
\label{fig:DIS}
\end{figure}

How should we calculate $\sigma(E,E_{\rm cut})$? Our process is quite analogous to deeply inelastic lepton scattering. We can take advantage of that. There is a standard analysis that allows us to write the cross section for $\chi$ scattering from the nucleus via vector boson exchange in terms of two structure functions, $F_\LT$ and $F_\LL$. In this section, we apply this standard analysis to $\chi$ scattering, using variables that are convenient for our present purposes. Although this analysis substantially simplifies the problem, it does not tell us what the structure functions $F_\LT$ and $F_\LL$ are. We will examine two rather different models for the structure functions in the following two sections.

The $\chi$ particle exchanges a virtual dark vector boson with the nucleus, as depicted in Fig.~\ref{fig:DIS}. The $\chi$ particle has momentum $p_\chi$ before the scattering and momentum $p'_\chi$ after the scattering. The dark vector boson carries spacelike momentum $q = p_\chi - p'_\chi$. One defines $Q^2 = - q^2$ so that $Q^2 > 0$. We define $\nu$ to be the energy of the vector boson in the nucleus rest frame. Thus the cut on the energy delivered to the nucleus is a cut $\nu > E_{\rm cut}$. We let $P$ be the momentum of the nucleus before the scattering and $M$ be its mass. Normally, $(P+q)^2 > M^2$, so that the scattering breaks up the nucleus. We define the Bjorken scaling variable $x_{\rm bj}$ by
\begin{equation}
x_{\rm bj} = \frac{Q^2}{2 M \nu}~.
\label{eq:bjx}
\end{equation}
We use the mass $M$ of the nucleus here. If we were to consider the nucleus as consisting of $A$ independent nucleons, then we might instead use $Ax_{\rm bj} = {Q^2}/(2 m_p \nu)$. 

Using lowest order perturbation theory in the interactions of the vector boson and using Lorentz invariance, parity invariance, and current conservation for the strong interactions, the differential cross section has the form familiar from deeply inelastic lepton scattering:
\begin{equation}
d\sigma = \frac{1}{4M[{E^2 - m_\chi^2}]^{1/2}}\,
(2\pi)^{-3}
d^4 p'_\chi\ \delta({p'_\chi}^2- m_\chi^2)\,
\frac{g^2_{\chi\bar \chi v}L^{\mu\nu}\ 4\pi g^2_{q\bar qv}W_{\mu\nu}}{(q^2 - m_v^2)^2}
\;,
\end{equation}
where $L^{\mu\nu}$ is 
\begin{equation}
L^{\mu\nu} = 4 p_\chi^\mu p_\chi^\nu 
- 2 (p_\chi^\mu q^\nu + q^\mu p_\chi^\nu) 
+ q^2 g^{\mu\nu}
\end{equation}
and $W_{\mu\nu}$ is the hadronic matrix element of the quark currents to which the vector particle couples, not including the coupling $g^2_{q\bar qv}$ but including a conventional factor $1/(4\pi)$,
\begin{equation}
W_{\mu\nu} = \frac{1}{4\pi} \sum_X
\bra{P} J_\mu(0) \ket{X}
\bra{X} J_\nu(0) \ket{P}
(2\pi)^4\delta(P + q - p_X)
\;.
\end{equation}
With the use of Eq.~(\ref{eq:jacobian}) in Appendix A, this is
\begin{equation}
d\sigma = \frac{g^2_{\chi\bar \chi v}\, g^2_{q\bar qv}}{16\pi M}\,
\frac{d\nu\, dQ^2}{E^2 - m_\chi^2}\,
\frac{L^{\mu\nu} W_{\mu\nu}}{(Q^2 + m_v^2)^2}
\;.
\end{equation}
We use $\nu$ and $Q^2$ as integration variables instead of the components of $p'_\chi$. The kinematics impose limits on $\nu$ and $Q^2$, which we derive in Appendix \ref{sec:kinematics}. Defining
\begin{equation}
\label{eq:muofnudef}
\mu^2(\nu)
=
\frac{m_\chi^2 \nu^2}
{[E(E-\nu) - m_\chi^2] + \sqrt{[E(E-\nu) - m_\chi^2]^2 - m_\chi^2\nu^2}}
\end{equation}
from Eq.~(\ref{eq:muofnudef0}), the limits are (Eqs.~(\ref{eq:nurange}), (\ref{eq:Qsqbounds}), and (\ref{eq:QSqlessthan2mnu}))
\begin{equation}
\begin{split}
\label{eq:limits}
E_{\rm cut} <{}& \nu < E - m_\chi
\;,
\\
2 \mu^2(\nu) <{}& Q^2 
<
4 [E (E-\nu)-m_\chi^2]
- 2 \mu^2(\nu)
\;,
\\
Q^2 <{}& 2 M \nu
\;.
\end{split}
\end{equation}

Now we can write $W_{\mu\nu}$ in terms of standard structure functions,
\begin{equation}
W^{\mu\nu} = C_\LT^{\mu\nu} F_\LT(x_{\rm bj},Q^2)
+ C_\LL^{\mu\nu} F_\LL(x_{\rm bj},Q^2)
\;,
\end{equation}
where
\begin{equation}
\begin{split}
C_\LT^{\mu\nu} ={}& 
-g^{\mu\nu}
+ \frac{q^\mu q^\nu}{q^2}
+ \frac{2x_{\rm bj}}{P\cdot q + 2 x_{\rm bj} M^2}
\left(P^\mu - \frac{P\cdot q}{q^2}\,q^\mu\right)
\left(P^\nu - \frac{P\cdot q}{q^2}\,q^\nu\right)
\;,
\\
C_\LL^{\mu\nu} ={}& 
\frac{1}{P\cdot q + 2 x_{\rm bj} M^2}
\left(P^\mu - \frac{P\cdot q}{q^2}\,q^\mu\right)
\left(P^\nu - \frac{P\cdot q}{q^2}\,q^\nu\right)
\;.
\end{split}
\end{equation}
Notice that $C_\LT^{\mu\nu}q_\nu = C_\LL^{\mu\nu}q_\nu = 0$ and that $C_\LT^{\mu\nu}a_\nu  = 0$ for any vector $a$ in the $P$-$q$ plane while  $C_\LL^{\mu\nu}a_\nu  = 0$ for any vector orthogonal to $P$ and $q$. Thus $C_\LT$ corresponds to the exchange of transversely polarized virtual vector bosons while $C_\LL$ corresponds to the exchange of longitudinally polarized virtual vector bosons. The structure functions $F_\LT$ and $F_\LL$ are related to the standard structure functions $F_1$ and $F_2$ by $F_\LT = F_1$ and $F_\LL = (1+2x_{\rm bj} M^2/P\cdot q) F_2 - 2  x_{\rm bj} F_1$.

We can thus write the cross section in terms of structure functions as
\begin{equation}
d\sigma = \frac{g^2_{\chi\chi v}\, g^2_{qqv}}{16\pi M}\,
\frac{d\nu\, dQ^2}{E^2 - m_\chi^2}\,
\frac{1}{(Q^2 + m_v^2)^2}
\left[
C_\LT^{\mu\nu} L_{\mu\nu}\,F_\LT(x_{\rm bj},Q^2)
+
C_\LL^{\mu\nu} L_{\mu\nu}\,F_\LL(x_{\rm bj},Q^2)
\right]
\;.
\end{equation}
One finds
\begin{equation}
\begin{split}
\label{eq:CTandCL}
C_\LT^{\mu\nu} L_{\mu\nu} ={}& 
\frac{Q^2 (2E-\nu)^2 }{\nu^2 + Q^2}
+ Q^2
- 4 m_\chi^2
\;,
\\
C_\LL^{\mu\nu} L_{\mu\nu} ={}& M\nu\,\frac{4 E(E-\nu) - Q^2}{\nu^2 + Q^2}
\;.
\end{split}
\end{equation}
Thus
\begin{equation}
\begin{split}
\label{eq:crosssection}
d\sigma ={}& \frac{g^2_{\chi\chi v}\, g^2_{qqv}}{16\pi}\,
\frac{d\nu\, dQ^2}{E^2 - m_\chi^2}\,
\frac{\nu}{(Q^2 + m_v^2)^2}
\Bigg\{
\left[
\frac{(2E-\nu)^2}{\nu^2 + Q^2}
+  \frac{Q^2 - 4 m_\chi^2}{Q^2}
\right]
2 x_{\rm bj} F_\LT(x_{\rm bj},Q^2)
\\&\quad
+
\frac{4E(E-\nu)-Q^2}{\nu^2 + Q^2}\,
F_\LL(x_{\rm bj},Q^2)
\Bigg\}
\;.
\end{split}
\end{equation}
The cross section that we want, $\sigma(E,E_{\rm cut})$, is then this $d\sigma$ integrated over $\nu > E_{\rm cut}$, taking into account the kinematic constraints (\ref{eq:limits}). This result is exact within the approximation of considering single vector boson exchange, but, of course, we need to be able to calculate $F_\LT$ and $F_\LL$. We explore this in the following two sections.

\section{DIS model}
\label{sec:DIS}
One way is to approach this as deeply inelastic scattering, as depicted in Fig.~\ref{fig:DIS}. The $\chi$ exchanges a virtual $V$ that is absorbed by a quark in the nucleus. If $Q^2$ is large, there is a short distance interaction in which the vector boson interacts with a quark or gluon in the nucleus. There are also long range interactions, both in the initial state and in the final state. For an inclusive cross section like that considered here, the final state interactions do not affect the cross section. The initial state interactions do affect the cross section, but they can be factored into parton distribution functions. The short distance interaction can be calculated perturbatively. Thus $F_\LT$ and $F_\LL$ are written as a convolution of parton distribution functions with the partonic structure functions $\hat F_\LT$ and $\hat F_\LL$. 

We will work at lowest order in perturbation theory for $\hat F_\LT$ and $\hat F_\LL$. At lowest order, the contributions from the gluon parton distribution function vanish for both $\hat F_\LL = 0$ and $\hat F_\LT$. For quarks at lowest order, $\hat F_\LL = 0$ and $\hat F_\LT$ is simply a delta function that sets the quark momentum fraction equal to $x_{\rm bj}$. (There would be a squared charge, $g^2_{q\bar q v}$, but we have already factored that out of the hadronic matrix element.) That is, $F_\LL = 0$ and
\begin{equation}
\label{eq:FTfromfq}
F_\LT = \frac{1}{2 x_{\rm bj}}\sum_q x_{\rm bj} f_{q/A}(x_{\rm bj},Q^2)
\;.
\end{equation}
Here we sum over flavors of quarks and antiquarks, $q = \Lu,\bar\Lu, \Ld, \bar\Ld, \Ls, \bar\Ls$ under our assumption that the mediator particle $v$ couples equally to all the flavors. (However, we have omitted charm and bottom quarks here since the corresponding parton distribution functions are small.) We have multiplied and divided by $x_{\rm bj}$ so that one factor is $x_{\rm bj} f_{q/A}(x_{\rm bj},Q^2)$, which is relatively insensitive to $x_{\rm bj}$ at small $x_{\rm bj}$. We note that the parton distributions here are the distributions in the nucleus A. The distribution of partons in a nucleus may be related approximately to the distribution of partons in a proton. For instance, if A is a nucleus with baryon number $A$ and charge $Z$ then
\begin{equation}
f_{\Lu/A}(x_{\rm bj},Q^2) dx_{\rm bj} \approx [Z f_{\Lu/p}(A x_{\rm bj},Q^2)
+ (A-Z) f_{\Ld/p}(A x_{\rm bj},Q^2) ] d(Ax_{\rm bj})
\;.
\end{equation}
That is
\begin{equation}
\label{eq:nuclearpdfs}
f_{\Lu/A}(x_{\rm bj},Q^2) \approx AZ f_{\Lu/p}(A x_{\rm bj},Q^2)
+ A(A-Z) f_{\Ld/p}(A x_{\rm bj},Q^2)
\;.
\end{equation}
Note that there are two factors of $A$ or $Z$ here. However, we use parton distribution functions for the nucleus provided at leading order
by Hirai-Kumano-Nagai (HKNlo) \cite{HKNpartons}, rather than this approximate formula. 

Thus in the DIS model we have
\begin{equation}
\begin{split}
\label{eq:crosssectionpartonmodel}
d\sigma ={}& \frac{g^2_{\chi\chi v}\, g^2_{qqv}}{16\pi}\,
\frac{d\nu\, dQ^2}{E^2 - m_\chi^2}\,
\frac{\nu}{(Q^2 + m_v^2)^2}
\left[
\frac{(2E-\nu)^2}{\nu^2 + Q^2}
+  \frac{Q^2 - 4 m_\chi^2}{Q^2}
\right]
\sum_q x_{\rm bj}
f_{q/A}(x_{\rm bj},Q^2)
\;.
\end{split}
\end{equation}
This approximation for the cross section should work well as long as $Q^2$ is large, say larger than a few ${\rm GeV}^2$. However, our numerical studies indicate that a good part of the cross section can come from the integration region in which $Q^2 < 1 \GeV^2$. For that region, we need another model.

\section{Saturation model}
\label{sec:SAT}

\begin{figure}
\centering
\includegraphics[width = 8 cm]{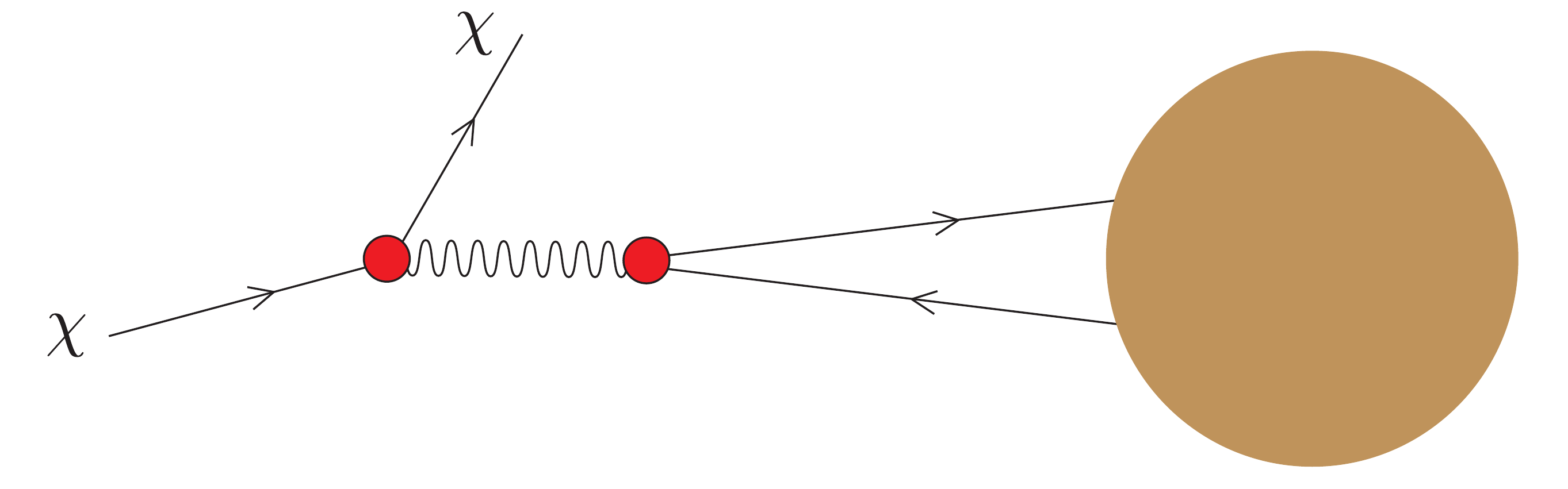}
\caption{Dipole picture for a $\chi$ particle scattering from a nucleus.}
\label{fig:Dipole}
\end{figure}

\begin{figure}
\centering
\includegraphics[width = 10 cm]{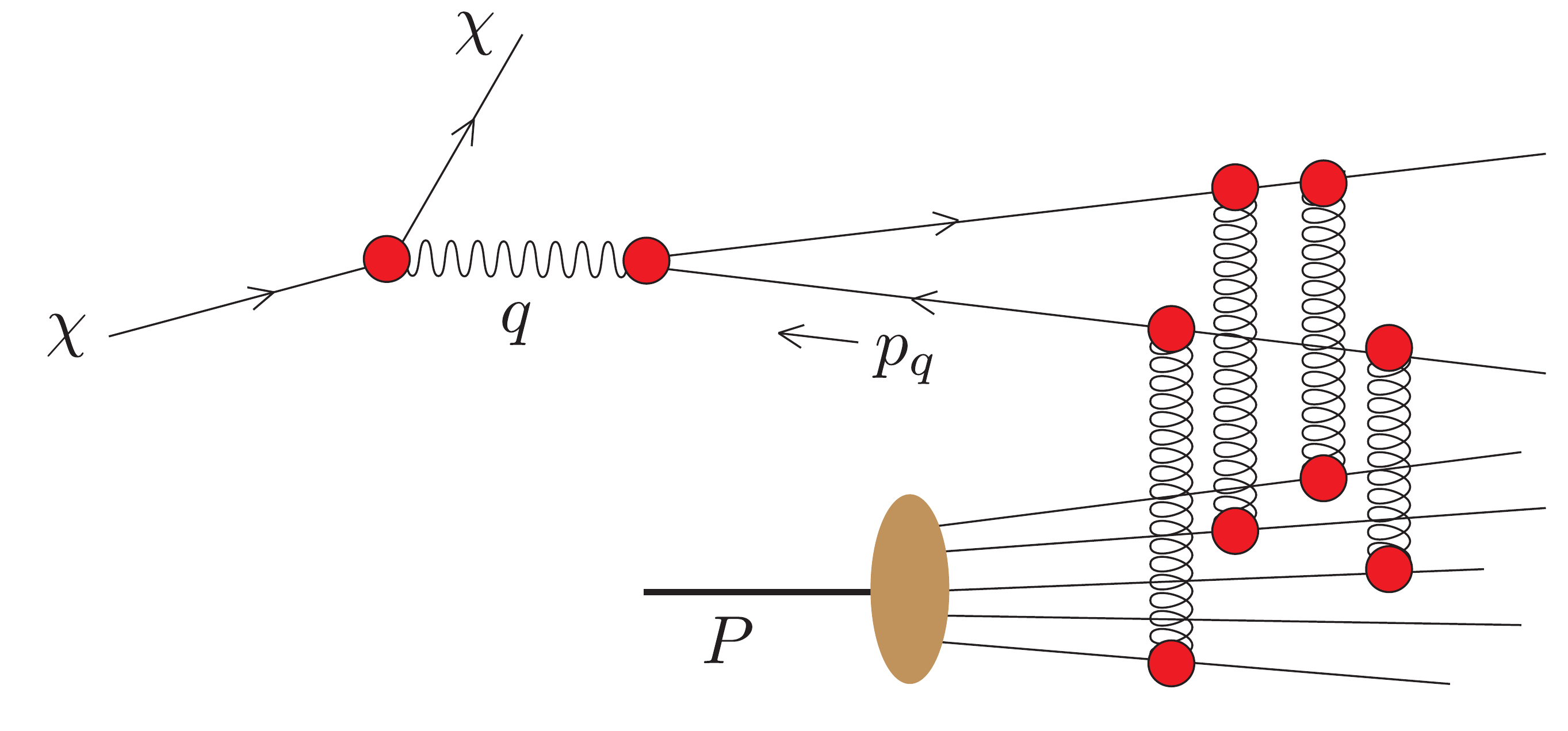}
\caption{The dipole created by the dark vector boson interacts with the nucleus via gluon exchange.}
\label{fig:Dipoledetails}
\end{figure}

There is another model available that should be useful for smaller values of $Q^2$ and large values of $\nu$. In this model, we view the interaction in the rest frame of the nucleus, as illustrated in Fig.~\ref{fig:Dipole}. The dark vector boson, carrying a large momentum, splits into a quark-antiquark pair. Each of the quark and antiquark also carry a large momentum as they move towards the nucleus. Thus they form a color dipole that can interact with the nucleus. The dipole interacts with the nucleus via gluon exchange, as illustrated in Fig.~\ref{fig:Dipoledetails}. We will model this interaction.

To motivate the model, it is helpful to examine the kinematics of the interaction in a little detail. We work in the rest frame of the nucleus and align the negative $z$-axis with the momentum $\vec q$ of the dark vector boson. Then, defining $q^\pm = (q^0 \pm q^3)/\sqrt 2$, we have $q^- \approx \sqrt 2 \nu$ and $q^+ \approx - 2^{-3/2}Q^2/\nu$. Thus in this frame $q^-$ is large and $q^+$ is small. In the Feynman diagram in Fig.~\ref{fig:Dipoledetails}, the dark vector boson couples to a quark propagator with momentum $p_q$, as in Fig.~\ref{fig:DIS}. We can estimate that $p_q^-$ is large while $p_q^+$ is small. Imagine writing the quark propagator in coordinate space, with the quark traveling through a space-time separation $\Delta x$ between the point where it interacts with a gluon from the nucleus and the point where it couples to the dark vector boson. Since $p_q \cdot \Delta x = p_q^+ \Delta x^- + p_q^- \Delta x^+ + p_q^\perp \cdot \Delta x^\perp$, we conclude that typically $\Delta x^-$ is large while $\Delta x^+$ is small. That is, the quark moves a long way in the minus direction. In fact, an estimate for $p_q^+$ is $p_q^+ \approx 2^{-3/2} Q^2/\nu$, so that an estimate for a typical range in the minus direction is $\Delta x^- = 2^{5/2} \pi\,\nu/Q^2$. Assuming that the first interaction of the quark with a gluon is inside the nucleus, this accounting puts the interaction of the quark with the dark vector boson well outside the nucleus when $\nu$ is large and $Q^2$ is not large.

This physical picture, depicted in Fig.~\ref{fig:Dipole}, seems at first to be completely different from the DIS picture of the previous section. Yet, if $\nu$ is very large and also $Q^2$ is large, both pictures can be correct and we can arrive at two ways of approximating the same cross section. The difference in the pictures arises from the difference of reference frames. The DIS picture is most easily derived in a reference frame in which the nucleus has a large momentum along the positive $z$-axis. The dipole picture of this cross section is most easily derived in the rest frame of the nucleus, with the dark vector boson having a large momentum along the negative $z$-axis.

We now need a model for $F_\LT$ and $F_\LL$ in the picture in which the dark vector boson turns into a quark-antiquark pair. The model, known as the saturation model, comes from the work of Nikolaev and Zakharov \cite{NikolaevZakharov},  Golec-Biernat and W\"usthoff \cite{GolecBiernatWusthoff1, GolecBiernatWusthoff2}, and  Mueller \cite{Mueller}. There is an extensive literature on the subject \cite{FrankfurtDipolesGluons, ForshawDipoles, McDermottDipoles, GotsmanReview, MuellerLecture, Bartels, GolecBiernat1, GolecBiernat2, HSdipole, Venugopalan, KowalskiDataAnalysis}. We will follow mostly Ref.~\cite{HSdipole} and will incorporate some refinements introduced by Bartels, Golec-Biernat, and Kowalski \cite{Bartels}.

When $Q^2$ is small, the longitudinal structure function $F_\LL$ is small compared to 
$2 x_{\rm bj} F_\LT$ since an on-shell massless vector boson does not have longitudinal 
polarizations. (For an analysis of $F_\LL$ in the saturation picture, see ref.~\cite{Machado:2006kd}.) 
Thus we simply approximate $F_\LL$ by zero in the saturation model, as we did in the DIS model. 
This leaves $F_\LT$. The result \cite{HSdipole} in the saturation model for $F_\LT$ is 
\begin{equation}
\label{eq:FTsaturation}
2 x_{\rm bj} F_\LT = \frac{1}{4\pi}\sum_f
\frac{24 Q^2}{(2\pi)^3}
\int\!d\bm b \int\!d\bm\Delta\
\frac{G(\sqrt{Q^2 + \Lambda_\rho^2}\,\Delta)}{\Delta^2}\,
\Xi(\bm b, \bm \Delta)
\;.
\end{equation}
Here one sums over quark flavors $f = \{u,d,s\}$ and the parameter $ \Lambda_\rho$ is discussed below. We integrate over a two dimensional vector $\bm b$ and a two dimensional vector $\bm \Delta$. The picture as outlined above is that the dark vector boson splits into a $q$-$\bar q$ pair, both with a large momentum in the direction of the dark vector boson momentum $q$. When this $q$-$\bar q$ pair reaches the nucleus, the quark is at transverse position $\bm b + \bm \Delta/2$ and the antiquark is a position $\bm b - \bm \Delta/2$. 

The function $G(\sqrt{Q^2 + \Lambda_\rho^2}\,\Delta)/\Delta^2$ represents the squared wave function for the $q$-$\bar q$ pair, integrated over the fraction $\alpha$ of the longitudinal momentum of the pair that is carried by the quark. The function $G(z)$ is
\begin{equation}
\label{eq:Gz}
G(z) = \int_0^1\!d\alpha\ [1-2\alpha(1-\alpha)]\,
\Big[\sqrt{\alpha(1-\alpha)}\, z\, K_1(\sqrt{\alpha(1-\alpha)} z)\Big]^2
\;.
\end{equation}
Here $K_1(x)$ is the modified Bessel function of order 1, equal to $-dK_0(x)/dx$. The function $G(z)$ equals 2/3 for $z = 0$. It behaves like $8/[3 z^2]$ for $z \to \infty$. Thus a rough approximation to it is
\begin{equation}
\label{eq:approxG}
G(z) \approx \frac{2}{3[1 + z^2/4]}
\;.
\end{equation}
This approximation is good to about 15\% for all values of $z$. 

We take the argument of $G$ to be $z = \sqrt{Q^2 + \Lambda_\rho^2}\,\Delta$. The perturbative calculation gives just $Q\Delta$. That means that the spatial extent of the wave function is of order $\Delta \sim 1/Q$. That should be right for large $Q$. But for small $Q$, we expect that the $q$ and $\bar q$ exchange gluons so as to bind themselves into one or more mesons -- predominantly a single $\rho$ meson. The $\rho$ meson has a size, which we can denote by $1/\Lambda_\rho$. To represent this non-perturbative effect, it seems sensible to replace $Q\Delta$ by $\sqrt{Q^2 + \Lambda_\rho^2}\,\Delta$. For the inverse radius of a $\rho$ meson, an approximate first guess might be $\Lambda_\rho \approx 1/(1\ \text{fm}) \approx 200 \MeV$.

The function $\Xi(\bm b, \bm \Delta)$ represents the probability that the $q$-$\bar q$ pair scatters from hadron $A$. If $\Delta$ is not small, then this probability is approximately 1 if either the quark or the antiquark hits hadron $A$. But if $\Delta$ is very small, the color dipole moment of the $q$-$\bar q$ pair is small and the pair can pass right through hadron $A$ without scattering. (This effect is known as {\it color transparency}). This suggests the following model (from Mueller \cite{Mueller} and Golec-Biernat and W\"usthoff \cite{GolecBiernatWusthoff1, GolecBiernatWusthoff2}).  We write\footnote{Golec-Biernat and W\"usthoff write this in the form $2\int\! d\bm b\  \Xi(\bm b, \bm \Delta) = \sigma_0 [1 - \exp(- \Delta^2/(2 R_0^2)]$, which is approximately equivalent when $\sigma_0$ and $R_0$ are suitably adjusted.}
\begin{equation}
\label{eq:Xi}
\Xi(\bm b, \bm \Delta) = 1 - e^{-\Delta^2 Q_\Ls^2(b)/4}
\;,
\end{equation}
where $Q_\Ls^2$ is the {\it saturation scale}. Evidently if $\Delta^2\ll 1/Q_\Ls^2$ then $\Xi(\bm b, \bm \Delta) \propto \Delta^2$ and the scattering probability tends to zero as $\Delta^2$ decreases. There is no scattering because the gluon field in hadron $A$ does not see the $q$-$\bar q$ pair. 

Before we go on to talk about the saturation scale $Q_\Ls^2(b)$, we should discuss Eq.~(\ref{eq:Xi}) and its connection to unitarity and to classical optics. Define $T(\bm b,\bm \Delta)$ by $\Xi(\bm b,\bm \Delta) = 1 - T(\bm b,\bm \Delta)$. We think of $\Xi$ as the probability for the dipole to be absorbed by the nucleus and we think of $T$ as the analogue of the transmission coefficient in optics \cite{HSdipole}. Let $R_A$ be the radius of the nucleus. We can then determine the necessary limiting properties of the function $T(\bm b,\bm \Delta)$. Here we follow Ref.~\cite{HSdipole}, which contains more details.
\begin{itemize}

\item If the dipole misses the nucleus, i.e. $|{\bm b}| > R_{A} + \Delta/2$, then $\Xi(\bm b,\bm \Delta)$ must be zero, therefore $T(\bm b,\bm \Delta) = 1$.

\item If the quark and the antiquark that make up the dipole are separated from each other by zero distance then, since it is a color singlet object, it simply passes through the nucleus. Therefore, $T(\bm b,\bm \Delta)=1$ for $\Delta=0$.

\item For small $\Delta$, the probability for the dipole to interact with the nucleus should be proportional to the square of the color dipole moment of the dipole: $T(\bm b,\bm \Delta) \propto \Delta^2$. We need $\Delta^2$ here because in the cut Feynman diagram for the process the dipole must exchange at least two gluons with the nucleus.

\item  For small $\Delta$, we can calculate the coefficient of $\Delta^2$ in $T$ using QCD perturbation theory.
 
\item $T(\bm b,\bm \Delta)\approx 0$ for large dipoles (large $\Delta$), when $|\bm b| < R_{A}$. That is, a large, strongly interacting dipole cannot pass through the nucleus leaving it intact.

\end{itemize}

To calculate the coefficient of $\Delta^2$ in $T$, we recognize that the probability that the gluon field does see the $q$-$\bar q$ pair depends not only on how small the color dipole moment is but also on how strong the gluon field is. Thus it is not surprizing that the saturation scale $Q_\Ls^2(b)$ in Eq.~(\ref{eq:Xi}) is proportional to the density of gluons in the nucleus:
\begin{equation}
\label{eq:Qs}
Q_\Ls^2(b) = \frac{2\pi^2 \as(\mu^2)}{3}\,
x G(x,\mu^2)\,\phi(b)
\;.
\end{equation}
Here $\phi(b)$ is modeled as a geometrical quantity that tells how the gluons are spread in the transverse separation from the center of  the nucleus:
\begin{equation}
\label{eq:phi}
\phi(b) = \frac{3}{2\pi R_A^3}\,\sqrt{R_A^2 - b^2}\,
\Theta(b^2 < R_A^2)
\;.
\end{equation}
The function $\phi(b)$ is normalized to $\int\!d\bm b\, \phi(b) = 1$. The function $G(x,\mu^2)$ is the gluon distribution function in the nucleus. We again employ the HKNlo distribution for lead, which is defined such that the total gluon distribution for the nucleus is given by $G(x,\mu^2) = A\, G_{\rm HKN}(A\,x_{\rm bj},\mu^2)$, which we insert in place of $G(x,\mu^2)$ in equation \eqref{eq:Qs}.

We need to set $\mu^2$ in $\as(\mu^2)$ and $x G(x,\mu^2)$ and we need to set $x$ in $x G(x,\mu^2)$. We follow the form of the choices of Bartels, Golec-Biernat, and Kowalski \cite{Bartels}. For the scale $\mu^2$, we take
\begin{equation}
\label{eq:BartelsMusq}
\mu^2 = \frac{C}{\Delta^2} + \mu_0^2
\;.
\end{equation}
The choice of a constant divided by $\Delta^2$ is sensible in the perturbative regime of small $\Delta^2$. However, for large $\Delta^2$ we do not want $\mu^2$ to be arbitrarily small. Thus we add a constant, $\mu_0^2$ to $C/\Delta^2$. We find a reasonable fit for $C = 6.00$ and $\mu_0^2 = 2.0 \GeV^2$. For the momentum fraction variable in the gluon distribution, we take
\begin{equation}
\label{eq:BartelsX}
x = \frac{Q^2 + 4 m_\Lq^2}{2 M \nu}
\;.
\end{equation}
This is $x_{\rm bj}$ when $Q^2$ is not too small. But for very small $Q^2$, we do not want $x$ to be arbitrarily small. Thus we add a small mass term, $4 m_\Lq^2$, to $Q^2$. This is in the same spirit as our adjustment of the argument of $G(z)$ in Eq.~(\ref{eq:FTsaturation}). Following Ref.~\cite{Bartels}, we take $m_\Lq = 140 \MeV$.

\begin{figure}
\centering
\includegraphics[width=10 cm]{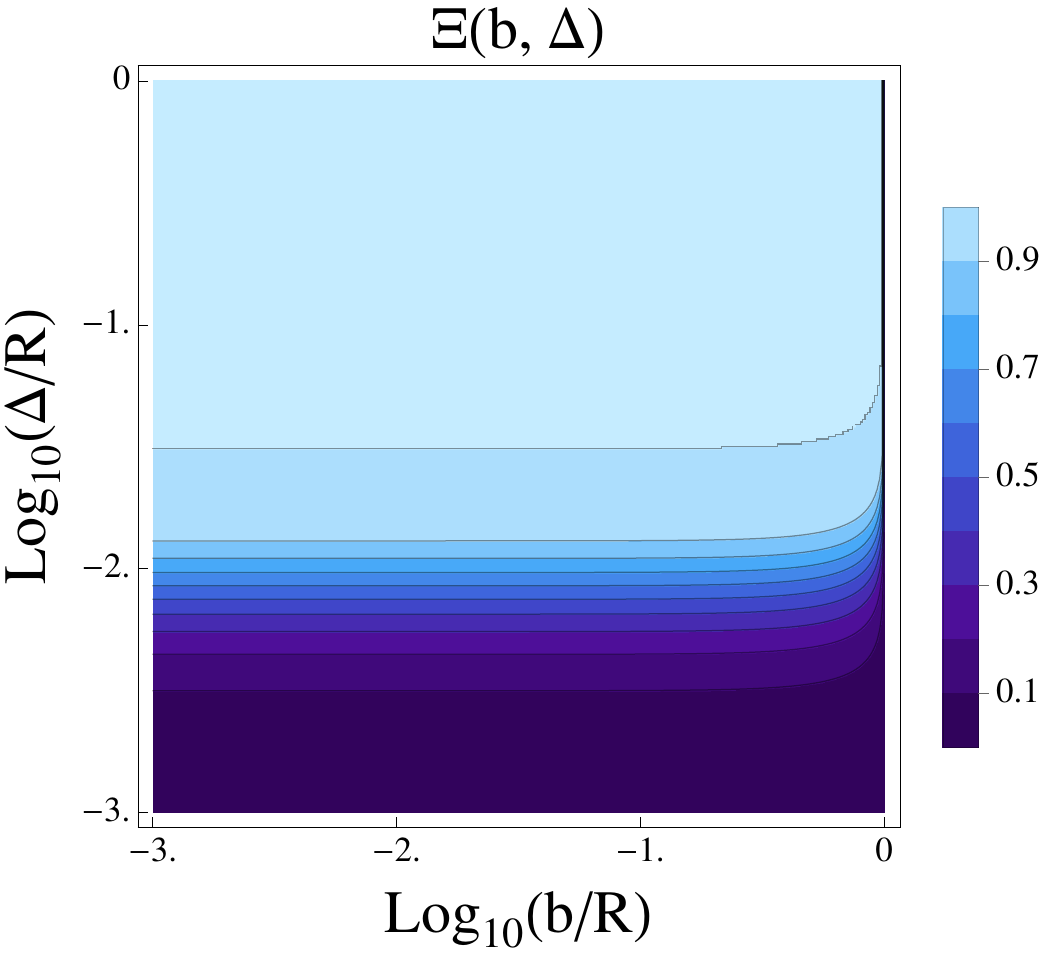}\\
\caption{The function $\Xi(\bm b, \bm \Delta)$ as a function of $|\bm b|/R$ and $|\bm \Delta|/R$ where $R$ is the radius of the lead nucleus. We calculate $\Xi(\bm b, \bm \Delta)$ using Eqs.~(\ref{eq:Xi}), (\ref{eq:Qs}), and (\ref{eq:phi}) with parameters given in Eqs.~(\ref{eq:BartelsMusq}) and  (\ref{eq:BartelsX}) and using HKNlo parton distributions for the distribution of gluons in a lead nucleus, with $x = 10^{-4}$ and $Q^2 = 1 \GeV^2$.}
\label{fig:Xi}
\end{figure}

We see that there is some QCD theory and some modeling in the net formula for $\Xi$. The resulting function $\Xi(\bm b, \bm \Delta)$ is illustrated in Fig.~\ref{fig:Xi}. We can perhaps appreciate from the figure that the model dependence is less than one might have thought. For $|\bm \Delta| > R/10$, $\Xi(\bm b, \bm \Delta)$ is very close to 1 for $|\bm b| < R$. When we get to $|\bm b| \approx R$, $\Xi$ drops very quickly to zero. The value $\Xi \approx 1$ is nonperturbative, but it is not really model dependent because 1 is the largest that $\Xi$ could be. For $|\bm \Delta| < R/10$, the behavior of $\Xi(\bm b, \bm \Delta)$ is not so trivial. However, this region is perturbative, so we have some control over the theory. In part, the shape is determined by the function $\phi(b)$ from Eq.~(\ref{eq:phi}). This part of the formula for $\Xi$ is simply a model for the distribution of gluons. The model is that the density of gluons is uniform throughout the nucleus. Thus there is some model dependence, but the model dependence is not too large.

There is more model dependence in the function $G(\sqrt{Q^2 + \Lambda_\rho^2}\,\Delta)/\Delta^2$ in Eq.~(\ref{eq:Gz}). This function is calculated using lowest order perturbation theory, so it should be accurate for large $Q^2$ and, correspondingly, small $\Delta$. For small $Q^2$ it simply represents a plausible model.

\section{Connection between the DIS and saturation models}
\label{sec:connection}
In Eq.~(\ref{eq:FTsaturation}), we can try to take the large $Q^2$ limit of $x_{\rm bj} F_\LT$ by taking the large $Q^2$ limit under the integration over $\bm\Delta$. In this limit, the argument, $\sqrt{Q^2 + \Lambda_\rho^2}\,\Delta$, of the function $G$ becomes just $Q\Delta$. Then for large $Q\Delta$ we have $G(Q\Delta) \sim 8/[3Q^2\Delta^2]$, as we noted earlier. We need to enforce that $Q\Delta$ is large inside the integration over $\bm \Delta$ and we do that in a crude way by inserting a factor $\Theta(\Delta > a/Q)$ for some constant $a$. This gives the approximation 

\begin{equation}
\label{eq:FTlargeQ}
2 x_{\rm bj} F_\LT \approx \sum_f
\frac{2}{\pi^4}
\int\!d\bm b \int\!d\bm\Delta\
\frac{\Theta(\Delta > a/Q)}{\Delta^4}\,
\Xi(\bm b, \bm \Delta)
\;.
\end{equation}
This matches with our DIS formula Eq.~(\ref{eq:FTfromfq}) if we identify
\begin{equation}
\label{eq:pdfdipole}
x\,f_{q/A}(x,Q^2) = 
\frac{1}{\pi^4}
\int\!d\bm b \int\!d\bm\Delta\
\frac{\Theta(\Delta > a/Q)}{\Delta^4}\,
\Xi(\bm b, \bm \Delta)
\;.
\end{equation}
There is a factor of 2 in this formula that results from suming over flavors $f$ in Eq.~(\ref{eq:FTlargeQ}) and over flavors and antiflavors in Eq.~(\ref{eq:FTfromfq}). The right hand side of this equation has some $x$ dependence because the gluon distribution that appears in the exponent in $\Xi$ depends on $x$. It is independent of the choice of quark flavor or antiflavor $q \in \{\Lu,\bar \Lu, \Ld,\bar \Ld, \Ls, \bar \Ls\}$.

There is a more direct approach to this, which was obtained in Ref.~\cite{HSdipole}. One starts directly with the operator definition of the parton distribution functions, $f_{q/A}(x,\mu^2)$, and analyzes the operator matrix element using the dipole picture. The operator matrix element requires ultraviolet renormalization, to eliminate a divergence from small $\Delta$ in the  integration over $\bm \Delta$. To match the standard  $\overline {\rm MS}$ definition of parton distribution functions, one should use dimensional regularization and an appropriate pole subtraction. However, one can obtain the same result at one loop order with a simple cut. The result of this analysis is Eq.~(\ref{eq:pdfdipole}) with
\begin{equation}
a = 2 e^{1/6 - \gamma_\LE} \approx 1.32657
\;.
\end{equation}

\begin{figure}
\centering
\includegraphics[width=14 cm]{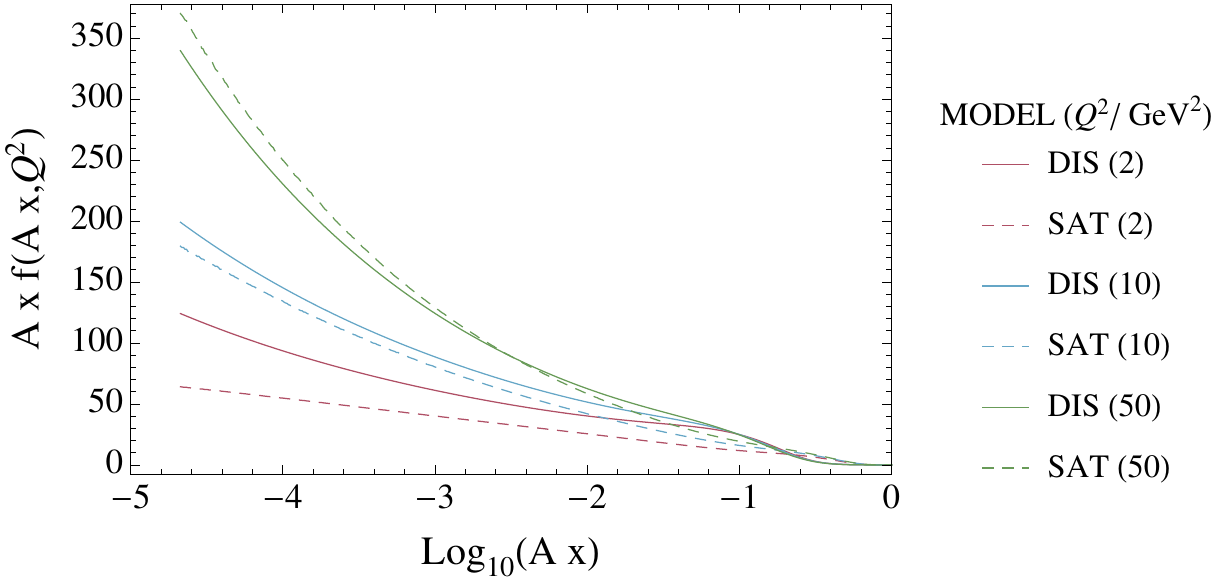}\\
\caption{The parton distribution function $f_{q/A}(x,Q^2)$ for $\bar {\rm u}$ quarks in a uranium nucleus according to the HKNlo parton distributions \cite{HKNpartons} used in this paper compared to the same distribution in the saturation model, Eq.~(\ref{eq:pdfdipole}). We plot $A x f_{q/A}(A x,Q^2)$ versus $\log_{10}(Ax)$ for $Q^2 =  2 \GeV^2$, $10 \GeV^2$, and $50 \GeV^2$.}
\label{fig:pdfCompare}
\end{figure}

Eq.~(\ref{eq:pdfdipole}) is based on lowest order perturbation theory for the wave function of the quark dipole, so one expects that it should begin to be accurate for $Q^2$ large enough so that perturbation theory applies. However the formula does not properly account for DGLAP evolution, so the result should begin to fail for very large $Q^2$. In Fig.~\ref{fig:pdfCompare}, we test how well this relationship works by ploting $A x f_{q/A}(A x,Q^2)$ versus $\log_{10}(Ax)$ for a few values of $Q^2$. We see that the approximation in Eq.~(\ref{eq:pdfdipole}) is only moderately successful at $Q^2 =  2 \GeV^2$, but that it works quite well for $Q^2 =  10 \GeV^2$. By $Q^2 =  50 \GeV^2$, it is still working quite well but is beginning to fail.

\section{Application to Scattering of Dark Matter}
\label{sec:dm}
We have studied the scattering of dark Dirac fermions through a vector mediated interaction with quarks. This amounts to a neutrino--like neutral current event, with the added theoretical interest of having no heavy electroweak boson to regulate the momentum transfer of the interaction. In this section we shall apply this formalism to a model of dark matter.

Continuing from Section \ref{sec:model}, with the scattering cross sections now in hand, it is straightforward to calculate the number of events expected in the detector. We first calculate the mean free path of the propagating dark particle,
\begin{equation}
  \lambda = \frac{1}{\rho_{A}\,\sigma(\chi N \to \chi N)},
\end{equation}
where $\rho_{A}$ is the number density of nuclei and $\sigma(\chi N \to \chi N)$ is the nuclear scattering cross section. The mean free path enters into the rescattering probability,
\begin{equation}
  P = \int^{L}_{0} dx\,\frac{1}{\lambda} e^{-\frac{x}{\lambda}} = 1 - e^{-\frac{L}{\lambda}}.
\end{equation}
The final number of events expected in the detector is,
\begin{equation}
N_{\rm det.} = \int\!dE\ (1 - P_{\rm shielding}(E))\times P_{\rm detector}(E) \times \frac{dN}{dE},
\label{eq:Nevents}
\end{equation}
where $dN/dE$ is defined in equation \eqref{eq:dNdE}. For scattering in the shielded region, composed of $\sim 15$~m of iron, we impose an arbitrary 1 GeV cut on the required energy transfer to prevent divergence of the deep inelastic cross section. We note that in practice, for the small values of the couplings that we can constrain, the probability of rescattering in shielding is extremely small, such that practically no scattering occurs. Further, since the probability of any given dark particle scattering is so low, one does not need to account for the degradation of the beam along
the length of the detector, and can approximate the scattering probability as simply $P \sim L/\lambda$. A fully realistic treatment would include multiple rescatterings, including low energy scatters that degrade the energy of incoming particles. This is not necessary for our purposes, which are adequately modeled by a single scattering event per dark particle. 

Using equation \eqref{eq:Nevents} and data provided by the experimental collaboration~\cite{Romanowski:1985xn,Duffy:1988rw} (and interpreted as below in~\cite{Golowich}), we may constrain our model. The E613 experiment delivered $1.8 \times 10^{17}$ protons on target (POT), and estimate that at most 100 detected events per $10^{17}$ POT represent muonless neutral current events at 90\% C.L. Thus, we exclude couplings where the number of expected detector events, $N_{\chi} > 180$.

\begin{figure}
\centering
\includegraphics[width=12cm]{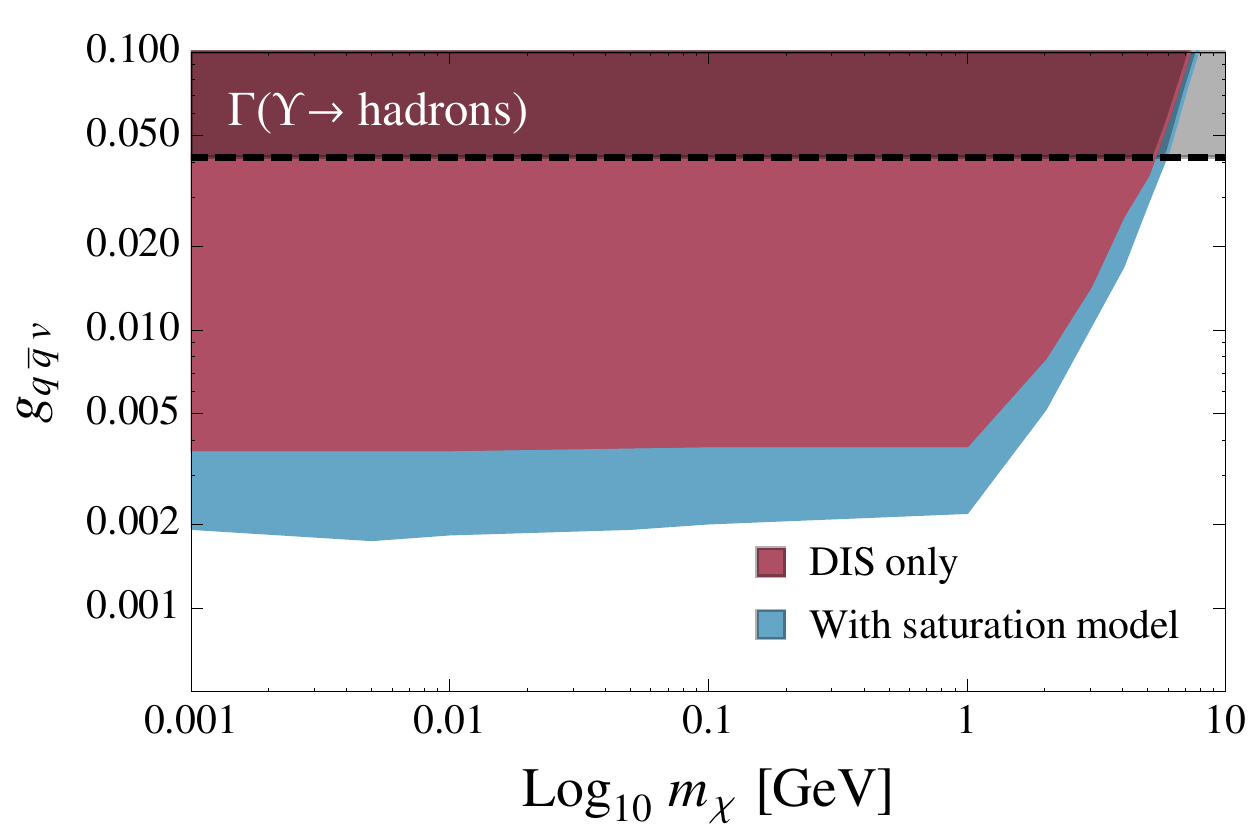}
\caption{Exclusion limits for the leptophobic
model described in the text, with $g_{\chi\bar{\chi}v} = 1$ and a mediator mass of 1 MeV. 
Also plotted is the region excluded by the study of $\Upsilon$ decays from~\cite{Carone:1994aa}.}
\label{fig:leptophobic}
\end{figure}

The result of this analysis as applied to the model (\ref{eq:lag}) is shown 
in Fig.~\ref{fig:leptophobic}, with the mediator mass set to $1 \MeV$ and the 
mediator--dark particle coupling fixed to unity. 
The two colored regions in the plot correspond to the scattering models described in 
Sections~\ref{sec:DIS} and~\ref{sec:SAT}. The ``DIS only'' region cuts off integration of the cross section 
for $Q^2 < 1 \GeV^2$, applying the scattering picture of Section~\ref{sec:DIS}. The region 
labeled ``With saturation model'' applies the same formalism, but additionally includes the dipole 
scattering mechanism described in Section~\ref{sec:SAT} for the $Q^2 < 1 \GeV^2$ region, resulting in 
a substantial improvement of the constraint. 
Also plotted is a mapping of the constraint on a leptophobic U(1) gauge boson, 
which couples to baryon number. Several constraints on such a model are 
described in \cite{Carone:1994aa}, the strongest of which (plotted) arises from the contribution of the new 
boson to the decay width of $\Upsilon$ mesons into hadronic final states. 
While we emphasize that this is a toy model, very similar models are of considerable 
phenomenological interest, and apt to be studied at existing fixed target facilities \cite{Batell:2014yra}. 

As another concrete example to demonstrate the impact of our formalism, we consider a ``minicharged'' particle scenario~\cite{Holdom:1985ag}, which is realized as the limit in which the mediator is a massless $U(1)$ vector boson which mixes kinetically with hypercharge~\cite{Okun:1982xi,Galison:1983pa,Holdom:1985ag,Dienes:1996zr,Jaeckel:2013ija}. The dark sector matter (the Dirac fermion) that is charged under the additional U(1) interacts with the standard model only through this mixing, which is parametrized via the mixing angle, $\kappa$, in the gauge invariant Lagrangian term $\mathcal{L} \supset -\frac{\kappa}{2}F^{\mu\nu}X_{\mu\nu}$, where $F^{\mu\nu}$ and $X^{\mu\nu}$ are respectively the field strength tensors of the SM and dark U(1) gauge groups.

\begin{figure}
\centering
\includegraphics[width=12 cm]{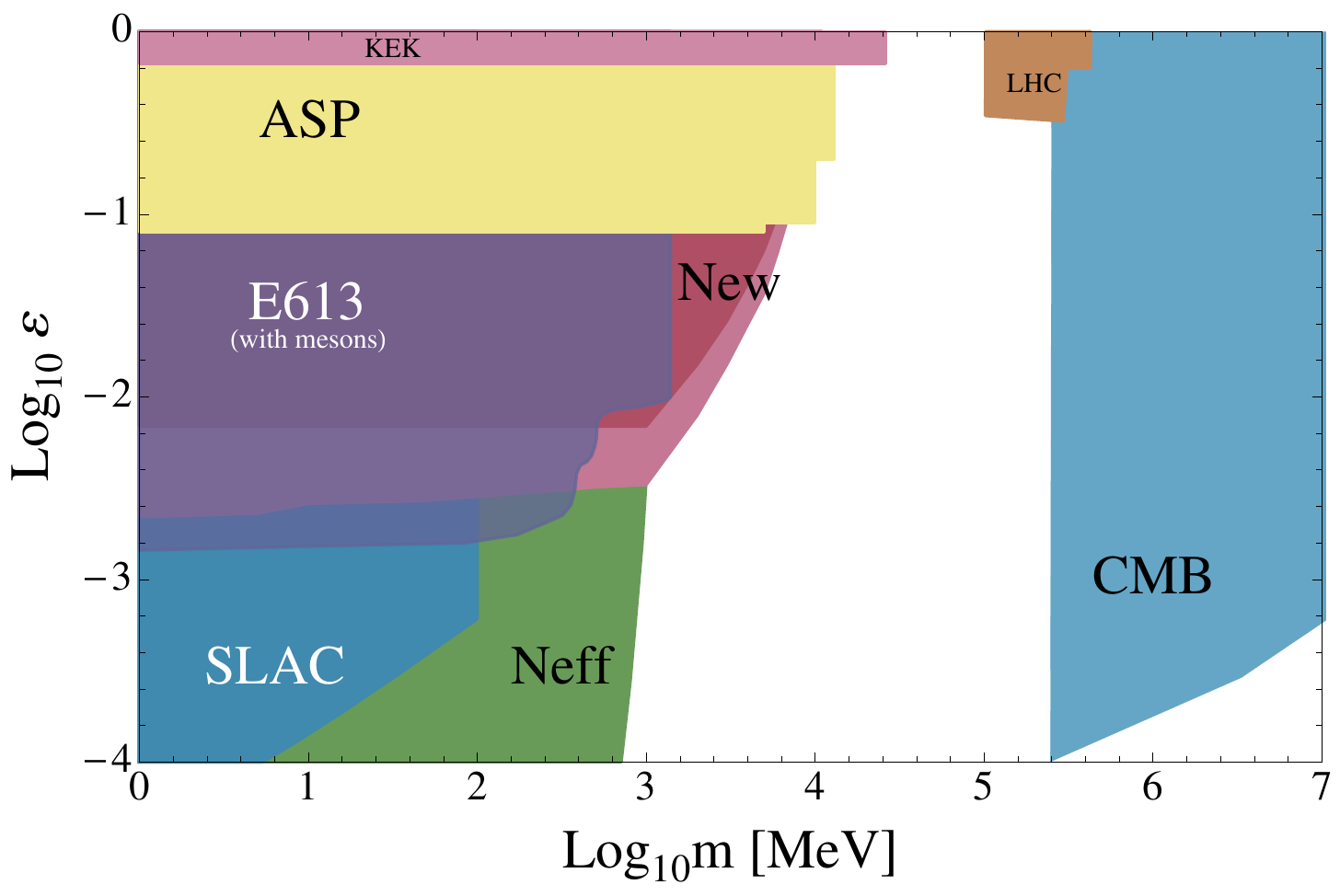}
\caption{Exclusion limits for minicharged particles in the MeV to GeV mass regime, including the results of this analysis. Other constraints are shown, arising from colliders \cite{Davidson:1991si}, a SLAC beam dump \cite{Prinz:1998ua}, the LHC \cite{Jaeckel:2012yz}, CMB \cite{Dolgov:2013una,Dubovsky:2003yn} and recent work on the number of light species, $N_{\rm eff}$ \cite{Vogel:2013raa}.}
\label{fig:excl_full}
\end{figure}

One can diagonalize the kinetic term in the Lagrangian with a field redefinition, the result of which is to induce electromagnetic interactions with the dark particles, which have an effective ``minicharge'', $\epsilon = \kappa g_{\rm h}/e$, where $g_{\rm h}$ is the hidden photon-dark fermion coupling and $e$ is the electromagnetic coupling constant. Then the cross section for both production and scattering scale with $\epsilon^2$. For our scenario involving quarks, the appropriate quark charges must be included in the cross section, such that the coupling of the mediator to the nucleus is correctly modeled as proceeding via mixing with the photon. The exclusion limits on the minicharge $\epsilon$ are shown in Fig.~\ref{fig:excl_full}.
It is worth noting that the constraint from the effective number of light particle species, $N_{\rm eff}$, is 
strong but subject to astrophysical uncertainties that make terrestrial collider based studies worthwhile.

\begin{figure}
\centering
\includegraphics[width=12 cm]{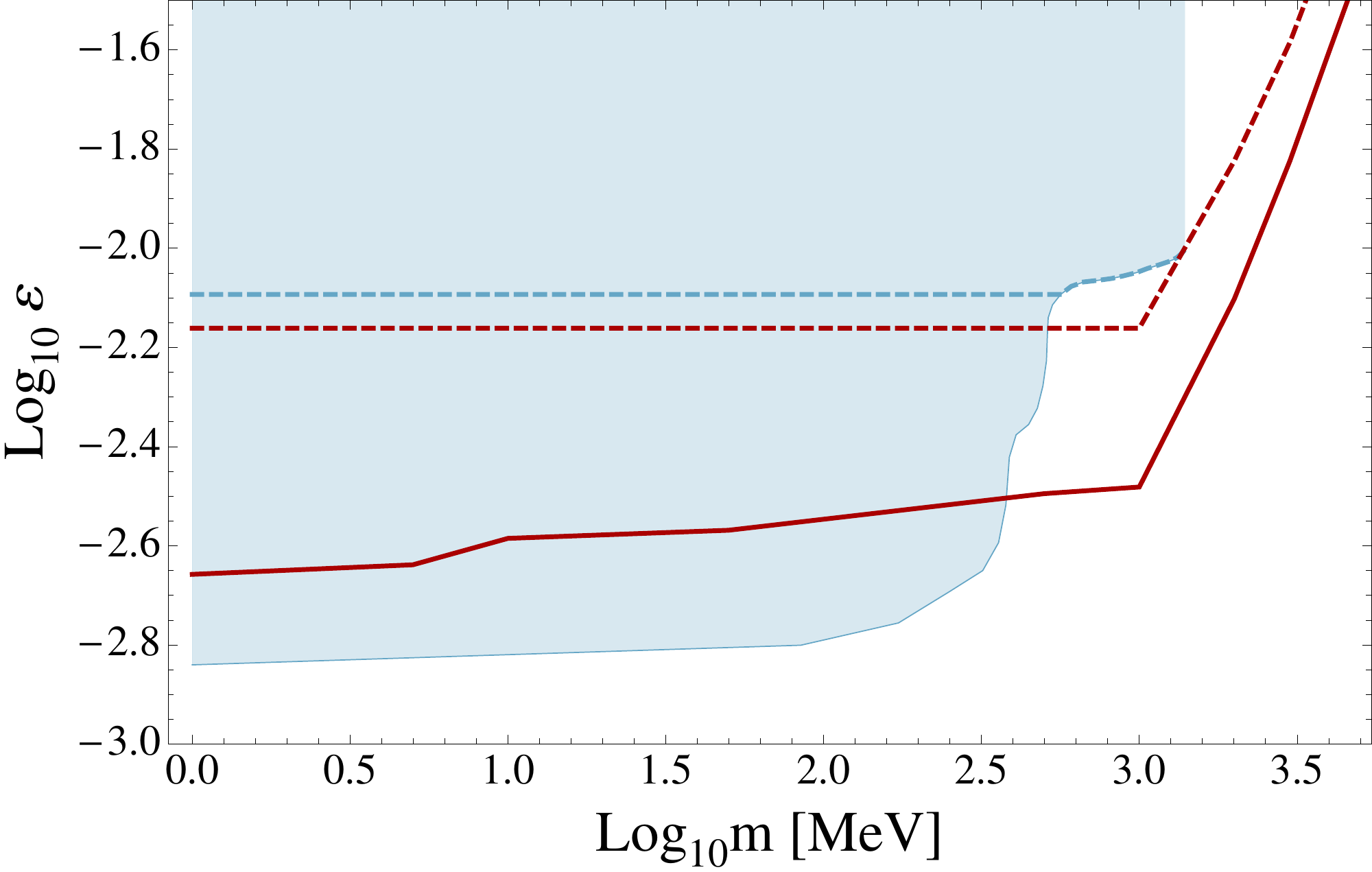}
\caption{The minicharge constraints arising only from the E613 experiment. See text for details.}
\label{fig:excl_e613}
\end{figure}

In Fig.~\ref{fig:excl_e613}, we compare (only) the constraints on the mini-charge model
previously derived from E613 \cite{Golowich} with those derived in this work, in the plane of the
mini-charged particle mass and $\epsilon$.  The results from our
analysis using only the deeply inelastic scattering regime are shown as the red dashed line,
whereas the inclusion of the low $Q^2 < 1\GeV^2$ regime via dipole scattering leads to the solid
red line.  A large improvement in the strength of the bound from the improved treatment of the low transfer
scattering is evident.  The previous constraint \cite{Golowich} is shown as the shaded region, and shows
a marked transition in the strength of the bound on $\epsilon$ by about
an order of magnitude as the particle mass crosses a few hundred MeV.  This sharp transition
is the result of dark particle production through meson decay, which switches off around 500 MeV,
leaving Drell-Yan production of the dark particles to dominate.  We have not included this production
mechanism in our bound, as it is model-dependent and
tangential to our goal of an improved description of the $\chi$-nucleus scattering cross section.
A more appropriate comparison of the impact of our improved computations is to the blue dashed
curve, which extrapolates the previous bound by extending the Drell-Yan-only limit to lower masses.
Of course, the actual bound on the mini-charged model at low mass
would be better represented by including the $\chi$ production from meson decay together with our improved
treatment of the scattering, though this is beyond the scope of this work.
Clearly, fixed target experiments are a fertile ground for testing this class of models.

\section{Conclusions}
\label{sec:conclusions}

We have investigated the detection of dark Dirac fermions in the context of the E613 beam dump experiment. The model employed gives rise to neutral current scattering, but in the absence of a heavy electroweak gauge boson to mediate the interaction. We studied the deep inelastic scattering in the detector in detail, and introduced a model valid at the low $Q^2$ values that become important in the absence of a heavy mediator to regulate the $1/Q^4$ behavior of the cross section. By including the effects of scattering at $Q^2 < 1 \GeV^2$ with a well theoretically motivated dipole model, we substantially improve upon constraints calculated using parton-level deep inelastic scattering alone. This could be especially relevant for new particle searches at future high energy beam dump facilities, which would allow access to regions of low $Q^2$ and small Bjorken-$x$.

\section*{Acknowledgements}

DES thanks Francesco Hautmann for helpful conversations. CJW thanks ITP Heidelberg for hospitality while performing some of this work, Joerg Jaeckel for providing questions and C\'eline B\oe hm for now ancient but very useful conversations. MS acknowledges the hospitality of the Aspen Center for Physics while part of this work was finished. MS and CJW thank Lucian Harland-Lang for enlightenment regarding nuclear parton distribution functions. All authors are grateful to Paolo Gondolo for the suggestion to consider extending an earlier analysis of MINOS to old beam dump experiments. The research of TMPT is supported in part by NSF grant PHY-1316792 and by the University of California, Irvine through a Chancellor's Fellowship. The research of DES was supported in part by the United States Department of Energy.

\bigskip\hrule\bigskip

\appendix
\section{Kinematics}
\label{sec:kinematics}

In this appendix, we present some of the details for the cross section for scattering a dark spin 1/2 particle, $\chi$, with momentum $p_\chi$, from a hadron A with momentum $P$. The hadron can be a nucleus. The dark particle has mass $m_\chi$ while the hadron has mass $M$. The dark particle mass may be of order 1 GeV or it may be smaller. The dark particle energy in the hadron rest frame, which we call $E$, is large compared to 1 GeV. We will introduce two different models for the scattering cross section.

In the hadron rest frame, we write the components of $P$ and $p_\chi$ as 
\begin{eqnarray}
P &=& (M,0,0,0) 
\;, \\
p_\chi &=&  (E,0,0,k) 
\;,
\end{eqnarray}
where $k = \sqrt{E^2 - m_\chi^2}$. The final state $\chi$ has 4-momentum
\begin{equation}
p_\chi' = (E-\nu, k' \sin\theta \cos\phi, k' \sin\theta \sin\phi, k'\cos\theta)
\;,
\end{equation}
where
\begin{equation}
\label{eq:kprimedef}
k' = \sqrt{(E-\nu)^2 - m_\chi^2}
\;.
\end{equation}
We suppose that the $\chi$ in the final state is not observed. Thus we will integrate over $p_\chi'$. 

The momentum transfer is
\begin{equation}
q = p_\chi - p_\chi'
\end{equation}
and is characterized by the energy transfer $\nu$, 
\begin{equation}
\label{eq:qdotP}
q \cdot P = M \nu
\;,
\end{equation}
and by the invariant
\begin{equation}
Q^2 = - q^2
\end{equation}
with $Q^2 > 0$.  In terms of the final state $\chi$ momentum,
\begin{equation}
Q^2 = 
2 E (E-\nu)
- 2 k k' \cos\theta
-2m_\chi^2 
\;.
\end{equation}
We define
\begin{equation}
x_{\rm bj} = \frac{Q^2}{2 M\nu}
\;.
\end{equation}
Note that $x_{\rm bj} \le 1$. Also note that when A is a nucleus of baryon number $A$, one often defines a scaled $x_{\rm bj}$ equal to $A Q^2/(2 M \nu)$. We do not do that here. However, we note that the ultimate limit on $x_{\rm bj}$ is $x_{\rm bj} \le 1$, but the practical limit beyond which the cross section is very small is $x_{\rm bj} \le 1/A$.

We will integrate over $Q^2$ and $\nu$ and will need the integration limits. Begin with $\nu$. We will impose a cut
\begin{equation}
\nu  > E_{\rm cut}
\;\;.
\end{equation}
That is, we wish to calculate the cross section for the process when at least a certain amount of energy $E_{\rm cut}$ is delivered to the hadron. Also  Eq.~(\ref{eq:kprimedef}) and $k^{\prime\,2} > 0$ gives $\nu < E - m_\chi$. Thus the integration range for $\nu$ is
\begin{equation}
\label{eq:nurange}
E_{\rm cut} < \nu < E - m_\chi
\;.
\end{equation}

Next, we need the limits on $Q^2$ at fixed $\nu$. Define a function $\mu^2(\nu)$ by
\begin{equation}
k k' = E (E-\nu) - \mu^2(\nu) - m_\chi^2
\;\;.
\end{equation}
Then
\begin{equation}
\label{eq:muofnudef0}
\mu^2(\nu)
=
\frac{m_\chi^2 \nu^2}
{[E(E-\nu) - m_\chi^2] + \sqrt{[E(E-\nu) - m_\chi^2]^2 - m_\chi^2\nu^2}}
\;.
\end{equation}
Then $Q^2$ has a simple form in terms of $\mu^2 (\nu )$,
\begin{equation}
Q^2 =
2 [E (E-\nu)-m_\chi^2] [1 - \cos\theta]
+ 2 \mu^2(\nu) \cos\theta
\;\;.
\end{equation}
One boundary of the integration region is forward scattering, $\cos\theta = 1$. 
On this boundary we have
\begin{equation}
Q^2 =
2 \mu^2(\nu)
\;\;,
\hskip 2 cm \cos\theta = 1
\;\;.
\end{equation}
The other boundary of the integration region is at $\cos\theta = -1$. There, we have
\begin{equation}
Q^2 =
4 [E (E-\nu)-m_\chi^2]
- 2 \mu^2(\nu)
\;\;,
\hskip 2 cm \cos\theta = -1
\;\;.
\end{equation}
For small $m_\chi$, this is $Q^2 \approx 4 E (E-\nu)$ with small corrections. Put together, the inequalities $-1 < \cos \theta < 1$ lead to
\begin{equation}
\label{eq:Qsqbounds}
2 \mu^2(\nu) < Q^2 
<
4 [E (E-\nu)-m_\chi^2]
- 2 \mu^2(\nu)
\;\;.
\end{equation}

There is a separate upper bound for $Q^2$. The momentum $q$ is absorbed by the hadron, giving a final state with momentum $P+q$. We need $(P+q)^2 > M^2$. This condition gives $x_{\rm bj} < 1$ or
\begin{equation}
\label{eq:QSqlessthan2mnu}
Q^2 < 2 M \nu
\;.
\end{equation}

Having found the integration limits, we translate the integration over $p_\chi'$ into integration over $\nu$ and $Q^2$ (integrating over the azimuthal angle $\phi$ to give a factor 
$2\pi$).  We obtain,
\begin{equation}
\begin{split}
\label{eq:jacobian}
d^4 p'_\chi\ \delta({p'_\chi}^2- m_\chi^2) ={}& \frac{k'^2 dk'}{2E'}\, d\cos\theta\, d\phi
\\
={}& \frac{\pi}{2\sqrt{E^2 - m_\chi^2}}\ d\nu\, dQ^2
\;\;.
\end{split}
\end{equation}
where $E' = E - \nu$ and $k'$ is given by Eq.~(\ref{eq:kprimedef}).

We will introduce two different models for the structure functions, and in particular for $F_\LT$. We will simply state the results of these models. However, if we want to examine the physics behind the models, it is convenient to use choose our reference frame wisely. We note that $F_\LT$ and $F_\LL$ depend only on $q$ and $P$. Thus we should choose a frame in which $q$ and $P$ are simple. We choose a frame in which both $q$ and $P$ have no transverse components and in which $q$ has a positive 3-component. Additionally, we now write momentum components in $(p^+,p^-,\bm p^T)$ format with $p^\pm = (p^0 \pm p^3)/\sqrt 2$. In our new frame,
\begin{equation}
\begin{split}
P ={}& (M/\sqrt 2,M/\sqrt 2, \bm 0)
\;\;,
\\
q ={}& \left(
\frac{1}{\sqrt 2}
\left[\nu + \sqrt{\nu^2 + Q^2}\right],
-\frac{1}{\sqrt 2}\,\frac{Q^2}{\nu + \sqrt{\nu^2 + Q^2}},
\bm 0
\right)
\end{split}
\end{equation}
In the kinematic region important for this paper, $Q^2 \ll \nu^2$, so that
\begin{equation}
q \approx \left(
\sqrt 2\, \nu,
-\frac{Q^2}{2\sqrt 2\, \nu},
 \bm 0
\right)
\;\;.
\end{equation}
Thus $q^+ \gg q^-$.


\newpage
\begin{thebibliography}{99}

%%% Intro:

%\cite{Bertone:2004pz}
\bibitem{Bertone:2004pz} 
G.~Bertone, D.~Hooper and J.~Silk,
%``Particle dark matter: Evidence, candidates and constraints,''
Phys.\ Rept.\ {\bf 405}, 279 (2005) [hep-ph/0404175].
%%CITATION = HEP-PH/0404175;%%
%1748 citations counted in INSPIRE as of 20 Jun 2014

%%% Direct Detection:

%\cite{Akerib:2013tjd}
\bibitem{Akerib:2013tjd} 
  D.~S.~Akerib {\it et al.}  [LUX Collaboration],
  %``First results from the LUX dark matter experiment at the Sanford Underground Research Facility,''
  arXiv:1310.8214 [astro-ph.CO].
  %%CITATION = ARXIV:1310.8214;%%
  %184 citations counted in INSPIRE as of 21 Apr 2014

%\cite{Aprile:2012nq}
\bibitem{Aprile:2012nq} 
  E.~Aprile {\it et al.}  [XENON100 Collaboration],
  %``Dark Matter Results from 225 Live Days of XENON100 Data,''
  arXiv:1207.5988 [astro-ph.CO].
  %%CITATION = ARXIV:1207.5988;%%
  
%\cite{Ahmed:2009rh}
\bibitem{Ahmed:2009rh} 
  Z.~Ahmed {\it et al.}  [CDMS Collaboration],
  %``Analysis of the low-energy electron-recoil spectrum of the CDMS experiment,''
  Phys.\ Rev.\ D {\bf 81}, 042002 (2010)
  [arXiv:0907.1438 [astro-ph.GA]].
  %%CITATION = ARXIV:0907.1438;%%
  
  %\cite{Angloher:2011uu}
\bibitem{Angloher:2011uu} 
  G.~Angloher, M.~Bauer, I.~Bavykina, A.~Bento, C.~Bucci, C.~Ciemniak, G.~Deuter and F.~von Feilitzsch {\it et al.},
  %``Results from 730 kg days of the CRESST-II Dark Matter Search,''
  Eur.\ Phys.\ J.\ C {\bf 72}, 1971 (2012)
  [arXiv:1109.0702 [astro-ph.CO]].
  %%CITATION = ARXIV:1109.0702;%%
  
\bibitem{Essig:2012yx} 
  R.~Essig, A.~Manalaysay, J.~Mardon, P.~Sorensen and T.~Volansky,
  %``First Direct Detection Limits on sub-GeV Dark Matter from XENON10,''
  Phys.\ Rev.\ Lett.\  {\bf 109}, 021301 (2012)
  [arXiv:1206.2644 [astro-ph.CO]].
  %%CITATION = ARXIV:1206.2644;%%
  
%\cite{Cushman:2013zza}
  \bibitem{Cushman:2013zza} 
  P.~Cushman, C.~Galbiati, D.~N.~McKinsey, H.~Robertson, T.~M.~P.~Tait, D.~Bauer, A.~Borgland and B.~Cabrera {\it et al.},
  %``Working Group Report: WIMP Dark Matter Direct Detection,''
  arXiv:1310.8327 [hep-ex].
  %%CITATION = ARXIV:1310.8327;%%
  %18 citations counted in INSPIRE as of 20 Jun 2014

%%% Colliders:

%\cite{Goodman:2010yf}
\bibitem{Goodman:2010yf} 
  J.~Goodman, M.~Ibe, A.~Rajaraman, W.~Shepherd, T.~M.~P.~Tait and H.~-B.~Yu,
  %``Constraints on Light Majorana dark Matter from Colliders,''
  Phys.\ Lett.\ B {\bf 695}, 185 (2011)
  [arXiv:1005.1286 [hep-ph]].
  %%CITATION = ARXIV:1005.1286;%%
  
%\cite{Bai:2010hh}
\bibitem{Bai:2010hh} 
  Y.~Bai, P.~J.~Fox and R.~Harnik,
  %``The Tevatron at the Frontier of Dark Matter Direct Detection,''
  JHEP {\bf 1012}, 048 (2010)
  [arXiv:1005.3797 [hep-ph]].
  %%CITATION = ARXIV:1005.3797;%%  
  
  %\cite{Goodman:2010ku}
\bibitem{Goodman:2010ku} 
  J.~Goodman, M.~Ibe, A.~Rajaraman, W.~Shepherd, T.~M.~P.~Tait and H.~-B.~Yu,
  %``Constraints on Dark Matter from Colliders,''
  Phys.\ Rev.\ D {\bf 82}, 116010 (2010)
  [arXiv:1008.1783 [hep-ph]].
  %%CITATION = ARXIV:1008.1783;%%
 
%\cite{Fox:2011pm}
\bibitem{Fox:2011pm} 
  P.~J.~Fox, R.~Harnik, J.~Kopp and Y.~Tsai,
  %``Missing Energy Signatures of Dark Matter at the LHC,''
  Phys.\ Rev.\ D {\bf 85}, 056011 (2012)
  [arXiv:1109.4398 [hep-ph]].
  %%CITATION = ARXIV:1109.4398;%%

  %\cite{Rajaraman:2011wf}
\bibitem{Rajaraman:2011wf} 
  A.~Rajaraman, W.~Shepherd, T.~M.~P.~Tait and A.~M.~Wijangco,
  %``LHC Bounds on Interactions of Dark Matter,''
  Phys.\ Rev.\ D {\bf 84}, 095013 (2011)
  [arXiv:1108.1196 [hep-ph]].
  %%CITATION = ARXIV:1108.1196;%%    
  
  %\cite{Bai:2012xg}
\bibitem{Bai:2012xg} 
  Y.~Bai and T.~M.~P.~Tait,
  %``Searches with Mono-Leptons,''
  arXiv:1208.4361 [hep-ph].
  %%CITATION = ARXIV:1208.4361;%%

  %%% Collider Exp Bounds
  
  %\cite{Aaltonen:2012jb}
\bibitem{Aaltonen:2012jb} 
  T.~Aaltonen {\it et al.}  [CDF Collaboration],
  %``A Search for dark matter in events with one jet and missing transverse energy in $p\bar{p}$ collisions at $\sqrt{s} = 1.96$ TeV,''
  Phys.\ Rev.\ Lett.\  {\bf 108}, 211804 (2012)
  [arXiv:1203.0742 [hep-ex]].
  %%CITATION = ARXIV:1203.0742;%%
  
  %\cite{Cheung:2012gi}
\bibitem{Cheung:2012gi} 
  K.~Cheung, P.~-Y.~Tseng, Y.~-L.~S.~Tsai and T.~-C.~Yuan,
  %``Global Constraints on Effective Dark Matter Interactions: Relic Density, Direct Detection, Indirect Detection, and Collider,''
  JCAP {\bf 1205}, 001 (2012)
  [arXiv:1201.3402 [hep-ph]].
  %%CITATION = ARXIV:1201.3402;%%    
  
  %\cite{Chatrchyan:2012tea}
\bibitem{Chatrchyan:2012tea} 
  S.~Chatrchyan {\it et al.}  [CMS Collaboration],
  %``Search for Dark Matter and Large Extra Dimensions in pp Collisions Yielding a Photon and Missing Transverse Energy,''
  arXiv:1204.0821 [hep-ex].
  %%CITATION = ARXIV:1204.0821;%%
  
  %\cite{Chatrchyan:2012pa}
\bibitem{Chatrchyan:2012pa} 
  S.~Chatrchyan {\it et al.}  [CMS Collaboration],
  %``Search for dark matter and large extra dimensions in monojet events in pp collisions at sqrt(s)= 7 TeV,''
  arXiv:1206.5663 [hep-ex].
  %%CITATION = ARXIV:1206.5663;%%

%\cite{ATLAS:2012ky}
  \bibitem{ATLAS:2012ky} 
  .~Aad {\it et al.} [ATLAS Collaboration],
  %``Search for dark matter candidates and large extra dimensions in events with a jet and missing transverse momentum with the ATLAS detector,''
  JHEP {\bf 1304}, 075 (2013)
  [arXiv:1210.4491 [hep-ex]].
  %%CITATION = ARXIV:1210.4491;%%
  %110 citations counted in INSPIRE as of 20 Jun 2014

%%% Secluded scenario %%%

%\cite{Batell:2009yf}
\bibitem{Batell:2009yf} 
  B.~Batell, M.~Pospelov and A.~Ritz,
  %``Probing a Secluded U(1) at B-factories,''
  Phys.\ Rev.\ D {\bf 79}, 115008 (2009)
  [arXiv:0903.0363 [hep-ph]].
  %%CITATION = ARXIV:0903.0363;%%
  %99 citations counted in INSPIRE as of 13 Dec 2013

  %\cite{Batell:2009di}
\bibitem{Batell:2009di} 
  B.~Batell, M.~Pospelov and A.~Ritz,
  %``Exploring Portals to a Hidden Sector Through Fixed Targets,''
  Phys.\ Rev.\ D {\bf 80}, 095024 (2009)
  [arXiv:0906.5614 [hep-ph]].
  %%CITATION = ARXIV:0906.5614;%%
  %86 citations counted in INSPIRE as of 13 Dec 2013

%\cite{Essig:2010gu}
\bibitem{Essig:2010gu} 
  R.~Essig, R.~Harnik, J.~Kaplan and N.~Toro,
  %``Discovering New Light States at Neutrino Experiments,''
  Phys.\ Rev.\ D {\bf 82}, 113008 (2010)
  [arXiv:1008.0636 [hep-ph]].
  %%CITATION = ARXIV:1008.0636;%%

%\cite{Hewett:2014qja}
  \bibitem{Hewett:2014qja} 
  J.~L.~Hewett, H.~Weerts, K.~S.~Babu, J.~Butler, B.~Casey, A.~de Gouvea, R.~Essig and Y.~Grossman {\it et al.},
  %``Planning the Future of U.S. Particle Physics (Snowmass 2013): Chapter 2: Intensity Frontier,''
  arXiv:1401.6077 [hep-ex].
  %%CITATION = ARXIV:1401.6077;%%

%%% Kinetic mixing
  
%\cite{Dienes:1996zr}
\bibitem{Dienes:1996zr} 
  K.~R.~Dienes, C.~F.~Kolda and J.~March-Russell,
  %``Kinetic mixing and the supersymmetric gauge hierarchy,''
  Nucl.\ Phys.\ B {\bf 492}, 104 (1997)
  [hep-ph/9610479];
  %%CITATION = HEP-PH/9610479;%%  
  %\cite{Rizzo:1998ut}
%\bibitem{Rizzo:1998ut} 
  T.~G.~Rizzo,
  %``Gauge kinetic mixing and leptophobic $Z^\prime$ in E(6) and SO(10),''
  Phys.\ Rev.\ D {\bf 59}, 015020 (1998)
  [hep-ph/9806397];
  %%CITATION = HEP-PH/9806397;%%
  %\cite{delAguila:1995rb}
%\bibitem{delAguila:1995rb} 
  F.~del Aguila, M.~Masip and M.~Perez-Victoria,
  %``Physical parameters and renormalization of U(1)-a x U(1)-b models,''
  Nucl.\ Phys.\ B {\bf 456}, 531 (1995)
  [hep-ph/9507455];
  %%CITATION = HEP-PH/9507455;%%
 %\cite{Kumar:2006gm}
%\bibitem{Kumar:2006gm} 
  J.~Kumar and J.~D.~Wells,
  %``CERN LHC and ILC probes of hidden-sector gauge bosons,''
  Phys.\ Rev.\ D {\bf 74}, 115017 (2006)
  [hep-ph/0606183].
  %%CITATION = HEP-PH/0606183;%% 

%\cite{Bjorken:2009mm}
\bibitem{Bjorken:2009mm} 
  J.~D.~Bjorken, R.~Essig, P.~Schuster and N.~Toro,
  %``New Fixed-Target Experiments to Search for Dark Gauge Forces,''
  Phys.\ Rev.\ D {\bf 80}, 075018 (2009)
  [arXiv:0906.0580 [hep-ph]].
  %%CITATION = ARXIV:0906.0580;%%

%\cite{Batell:2014mga}
  \bibitem{Batell:2014mga} 
  B.~Batell, R.~Essig and Z.~'e.~Surujon,
  %``Strong Constraints on Sub-GeV Dark Matter from SLAC Beam Dump E137,''
  arXiv:1406.2698 [hep-ph].
  %%CITATION = ARXIV:1406.2698;%%

%\cite{Carone:1994aa}
  \bibitem{Carone:1994aa} 
  C.~D.~Carone and H.~Murayama,
  %``Possible light U(1) gauge boson coupled to baryon number,''
  Phys.\ Rev.\ Lett.\ {\bf 74}, 3122 (1995)
  [hep-ph/9411256];
  %%CITATION = HEP-PH/9411256;%%
  %65 citations counted in INSPIRE as of 20 Jun 2014
  %\cite{Carone:1995pu}
  %\bibitem{Carone:1995pu} 
  C.~D.~Carone and H.~Murayama,
  %``Realistic models with a light U(1) gauge boson coupled to baryon number,''
  Phys.\ Rev.\ D {\bf 52}, 484 (1995)
  [hep-ph/9501220].
  %%CITATION = HEP-PH/9501220;%%
  %52 citations counted in INSPIRE as of 20 Jun 2014

%%% Light mediators at FT.

%\cite{deNiverville:2011it}
\bibitem{deNiverville:2011it} 
  P.~deNiverville, M.~Pospelov and A.~Ritz,
  %``Observing a light dark matter beam with neutrino experiments,''
  Phys.\ Rev.\ D {\bf 84}, 075020 (2011)
  [arXiv:1107.4580 [hep-ph]].
  %%CITATION = ARXIV:1107.4580;%%
  %29 citations counted in INSPIRE as of 22 Jun 2014

%\cite{deNiverville:2012ij}
\bibitem{deNiverville:2012ij} 
  P.~deNiverville, D.~McKeen and A.~Ritz,
  %``Signatures of sub-GeV dark matter beams at neutrino experiments,''
  Phys.\ Rev.\ D {\bf 86}, 035022 (2012)
  [arXiv:1205.3499 [hep-ph]].
  %%CITATION = ARXIV:1205.3499;%%
  %19 citations counted in INSPIRE as of 22 Jun 2014

%\cite{Batell:2014yra}
\bibitem{Batell:2014yra} 
  B.~Batell, P.~deNiverville, D.~McKeen, M.~Pospelov and A.~Ritz,
  %``Leptophobic Dark Matter at Neutrino Factories,''
  arXiv:1405.7049 [hep-ph].
  %%CITATION = ARXIV:1405.7049;%%
  %1 citations counted in INSPIRE as of 26 Jun 2014

%%% Tools %%%

%\cite{Alwall:2011uj}
\bibitem{Alwall:2011uj} 
  J.~Alwall, M.~Herquet, F.~Maltoni, O.~Mattelaer and T.~Stelzer,
  %``MadGraph 5 : Going Beyond,''
  JHEP {\bf 1106}, 128 (2011)
  [arXiv:1106.0522 [hep-ph]].
  %%CITATION = ARXIV:1106.0522;%%
  
%\cite{Christensen:2008py}
\bibitem{Christensen:2008py} 
  N.~D.~Christensen and C.~Duhr,
  %``FeynRules - Feynman rules made easy,''
  Comput.\ Phys.\ Commun.\  {\bf 180}, 1614 (2009)
  [arXiv:0806.4194 [hep-ph]].
  %%CITATION = ARXIV:0806.4194;%%

%%% E613 experimental %%%

%\cite{Romanowski:1985xn}
\bibitem{Romanowski:1985xn} 
  T.~A.~Romanowski,
  %``Prompt Neutrino Production In A 400-gev/c Proton Beam Dump Experiment,''
  Acta Phys.\ Polon.\ B {\bf 16}, 179 (1985).
  %%CITATION = APPOA,B16,179;%%

%\cite{Duffy:1988rw}
\bibitem{Duffy:1988rw} 
  M.~E.~Duffy, G.~K.~Fanourakis, R.~J.~Loveless, D.~D.~Reeder, E.~S.~Smith, S.~Childress, C.~Castoldi and G.~Conforto {\it et al.},
  %``Neutrino Production By 400-gev/c Protons In A Beam-dump Experiment,''
  Phys.\ Rev.\ D {\bf 38}, 2032 (1988).
  %%CITATION = PHRVA,D38,2032;%%

%%% Nuclear PDFs

\bibitem{HKNpartons}
  M.~Hirai, S.~Kumano and T.~-H.~Nagai,
  {\em Determination of nuclear parton distribution functions and 
  their uncertainties in next-to-leading order},
  Phys.\ Rev.\ C {\bf 76}, 065207 (2007)
  \href{http://inspirehep.net/record/761288?ln=en}{[INSPIRE]}.
  %%CITATION = ARXIV:0709.3038;%%

%%% Scattering:

\bibitem{NikolaevZakharov} 
  N.~N.~Nikolaev and B.~G.~Zakharov,
  {\em Color transparency and scaling properties of nuclear shadowing 
  in deep inelastic scattering},
  Z.\ Phys.\ C {\bf 49}, 607 (1991)
  \href{http://inspirehep.net/record/298382?ln=en}{[INSPIRE]}.
  %%CITATION = ZEPYA,C49,607;%%
      
\bibitem{GolecBiernatWusthoff1} 
  K.~J.~Golec-Biernat and M.~Wusthoff,
  {\em Saturation effects in deep inelastic scattering at low Q**2 
  and its implications on diffraction},
  Phys.\ Rev.\ D {\bf 59}, 014017 (1998)
  \href{http://inspirehep.net/record/473813?ln=en}{[INSPIRE]}.
    %%CITATION = HEP-PH/9807513;%%
  
\bibitem{GolecBiernatWusthoff2} 
  K.~J.~Golec-Biernat and M.~Wusthoff,
  {\em Saturation in diffractive deep inelastic scattering},
  Phys.\ Rev.\ D {\bf 60}, 114023 (1999)
  \href{http://inspirehep.net/record/496774?ln=en}{[INSPIRE]}.
  %%CITATION = HEP-PH/9903358;%%
  
\bibitem{Mueller} 
  A.~H.~Mueller,
  {\em Parton saturation at small x and in large nuclei},
  Nucl.\ Phys.\ B {\bf 558}, 285 (1999)
  \href{http://inspirehep.net/record/498572?ln=en}{[INSPIRE]}.
  %%CITATION = HEP-PH/9904404;%%
    
\bibitem{FrankfurtDipolesGluons} 
  L.~Frankfurt, A.~Radyushkin and M.~Strikman,
  {\em Interaction of small size wave packet with hadron target},
  Phys.\ Rev.\ D {\bf 55}, 98 (1997)
  \href{http://inspirehep.net/record/418564?ln=en}{[INSPIRE]}.
  %%CITATION = HEP-PH/9610274;%%

\bibitem{ForshawDipoles} 
  J.~R.~Forshaw, G.~Kerley and G.~Shaw,
  {\em Extracting the dipole cross-section from photoproduction 
  and electroproduction total cross-section data},
  Phys.\ Rev.\ D {\bf 60}, 074012 (1999)
  \href{http://inspirehep.net/record/496674?ln=en}{[INSPIRE]}.
  %%CITATION = HEP-PH/9903341;%%

\bibitem{McDermottDipoles} 
  M.~McDermott, L.~Frankfurt, V.~Guzey and M.~Strikman,
  {\em Unitarity and the QCD improved dipole picture},
  Eur.\ Phys.\ J.\ C {\bf 16}, 641 (2000)
  \href{http://inspirehep.net/record/512804?ln=en}{[INSPIRE]}.
  %%CITATION = HEP-PH/9912547;%%
  
\bibitem{GotsmanReview} 
  E.~Gotsman, E.~Levin, M.~Lublinsky, U.~Maor, E.~Naftali and K.~Tuchin,
  {\em Has HERA reached a new QCD regime?: (Summary of our view)},
  J.\ Phys.\ G {\bf 27}, 2297 (2001)
  \href{http://inspirehep.net/record/535295?ln=en}{[INSPIRE]}.
  %%CITATION = HEP-PH/0010198;%%

\bibitem{MuellerLecture}
  A.~H.~Mueller,
  {\em Parton saturation: An Overview},
  in J.P.Blaizot and E. Iancu, eds.\ 
  ``QCD perspectives on hot and dense matter. Proceedings, NATO Advanced Study Institute'',
  Cargese, France, August, 2001,
  hep-ph/0111244
  \href{http://inspirehep.net/record/609280}{[INSPIRE]}.
  %%CITATION = HEP-PH/0111244;%%
    
\bibitem{GolecBiernat1} 
  K.~J.~Golec-Biernat,
  {\em Physics of parton saturation},
  Acta Phys.\ Polon.\ B {\bf 35}, 3103 (2004)
  \href{http://inspirehep.net/record/669868?ln=en}{[INSPIRE]}.
  %%CITATION = APPOA,B35,3103;%%    

\bibitem{GolecBiernat2} 
  K.~J.~Golec-Biernat,
  {\em Theoretical review of diffractive phenomena},
  Nucl.\ Phys.\ A {\bf 755}, 133 (2005)
  \href{http://inspirehep.net/record/687954?ln=en}{[INSPIRE]}.
  %%CITATION = HEP-PH/0507251;%%
  
\bibitem{Venugopalan} 
  P.~Tribedy and R.~Venugopalan,
  {\em Saturation models of HERA DIS data and inclusive hadron distributions 
  in p+p collisions at the LHC},
  Nucl.\ Phys.\ A {\bf 850}, 136 (2011)
  [Erratum-ibid.\ A {\bf 859}, 185 (2011)]
   \href{http://inspirehep.net/record/875812?ln=en}{[INSPIRE]}.
  %%CITATION = ARXIV:1011.1895;%%

\bibitem{KowalskiDataAnalysis} 
  A.~Luszczak and H.~Kowalski,
  {\em Dipole model analysis of high precision HERA data},
  arXiv:1312.4060 [hep-ph]
  \href{http://inspirehep.net/record/1269467?ln=en}{[INSPIRE]}.
  %%CITATION = ARXIV:1312.4060;%%
 
\bibitem{HSdipole}
  F.~Hautmann, D.~E.~Soper,
  {\em Parton distribution function for quarks in an s-channel approach},
  Phys.\ Rev.\  {\bf D75}, 074020 (2007)
  \href{http://inspirehep.net/record/744097?ln=en}{[INSPIRE]}.
  %%CITATION = HEP-PH/0702077;%%

\bibitem{Bartels} 
  J.~Bartels, K.~J.~Golec-Biernat and H.~Kowalski,
  {\em A modification of the saturation model: DGLAP evolution},
  Phys.\ Rev.\ D {\bf 66}, 014001 (2002)
  \href{http://inspirehep.net/record/584724?ln=en}{[INSPIRE]}.
  %%CITATION = HEP-PH/0203258;%%

%%% F_L refs.

%\cite{Machado:2006kd}
\bibitem{Machado:2006kd}
  M.~V.~T.~Machado,
  %``Investigating F(L)(x,Q**2) at fixed energy in the color dipole formalism,''
  Eur.\ Phys.\ J.\ C {\bf 47} (2006) 365.
  %%CITATION = EPHJA,C47,365;%%
  %8 citations counted in INSPIRE as of 18 Oct 2013

%%% E613 References

% E613 minicharge constraints
%\cite{Golowich}
\bibitem{Golowich} 
  Golowich, E. and Robinett, R. W.
  %``Limits on millicharged matter from beam-dump experiments,''
  Phys.\ Rev.\ D {\bf 35}, 391 (1987).

%%% Hidden photon stuff

%\cite{Holdom:1985ag}
\bibitem{Holdom:1985ag} 
  B.~Holdom,
  %``Two U(1)'s and Epsilon Charge Shifts,''
  Phys.\ Lett.\ B {\bf 166}, 196 (1986).
  %%CITATION = PHLTA,B166,196;%%
  %713 citations counted in INSPIRE as of 12 Jun 2014

%\cite{Okun:1982xi}
\bibitem{Okun:1982xi} 
  L.~B.~Okun,
  %``Limits Of Electrodynamics: Paraphotons?,''
  Sov.\ Phys.\ JETP {\bf 56}, 502 (1982)
  [Zh.\ Eksp.\ Teor.\ Fiz.\  {\bf 83}, 892 (1982)].
  %%CITATION = SPHJA,56,502;%%
  %162 citations counted in INSPIRE as of 12 Jun 2014

%\cite{Galison:1983pa}
\bibitem{Galison:1983pa} 
  P.~Galison and A.~Manohar,
  %``TWO Z's OR NOT TWO Z's?,''
  Phys.\ Lett.\ B {\bf 136}, 279 (1984).
  %%CITATION = PHLTA,B136,279;%%
  %27 citations counted in INSPIRE as of 12 Jun 2014

%\cite{Jaeckel:2013ija}
\bibitem{Jaeckel:2013ija} 
  J.~Jaeckel,
  %``A force beyond the Standard Model - Status of the quest for hidden photons,''
  Frascati Phys.\ Ser.\  {\bf 56}, 172 (2012)
  [arXiv:1303.1821 [hep-ph]].
  %%CITATION = ARXIV:1303.1821;%%
  %19 citations counted in INSPIRE as of 12 Jun 2014

%%% Other minicharge constraints

%\cite{Davidson:1991si}
\bibitem{Davidson:1991si} 
  S.~Davidson, B.~Campbell and D.~C.~Bailey,
  %``Limits on particles of small electric charge,''
  Phys.\ Rev.\ D {\bf 43}, 2314 (1991).
  %%CITATION = PHRVA,D43,2314;%%
  %80 citations counted in INSPIRE as of 12 Jun 2014

%\cite{Prinz:1998ua}
\bibitem{Prinz:1998ua} 
  A.~A.~Prinz, R.~Baggs, J.~Ballam, S.~Ecklund, C.~Fertig, J.~A.~Jaros, K.~Kase and A.~Kulikov {\it et al.},
  %``Search for millicharged particles at SLAC,''
  Phys.\ Rev.\ Lett.\  {\bf 81}, 1175 (1998)
  [hep-ex/9804008].
  %%CITATION = HEP-EX/9804008;%%
  %53 citations counted in INSPIRE as of 12 Jun 2014

%\cite{Jaeckel:2012yz}
\bibitem{Jaeckel:2012yz}
  J.~Jaeckel, M.~Jankowiak and M.~Spannowsky,
  %``LHC probes the hidden sector,''
  Phys.\ Dark Univ.\  {\bf 2} (2013) 111
  [arXiv:1212.3620 [hep-ph]].
  %%CITATION = ARXIV:1212.3620;%%
  %12 citations counted in INSPIRE as of 12 Jun 2014

%\cite{Dolgov:2013una}
\bibitem{Dolgov:2013una} 
  A.~D.~Dolgov, S.~L.~Dubovsky, G.~I.~Rubtsov and I.~I.~Tkachev,
  %``Constraints on millicharged particles from Planck data,''
  Phys.\ Rev.\ D {\bf 88}, no. 11, 117701 (2013)
  [arXiv:1310.2376 [hep-ph]].
  %%CITATION = ARXIV:1310.2376;%%
  %11 citations counted in INSPIRE as of 12 Jun 2014

%\cite{Dubovsky:2003yn}
\bibitem{Dubovsky:2003yn} 
  S.~L.~Dubovsky, D.~S.~Gorbunov and G.~I.~Rubtsov,
  %``Narrowing the window for millicharged particles by CMB anisotropy,''
  JETP Lett.\  {\bf 79}, 1 (2004)
  [Pisma Zh.\ Eksp.\ Teor.\ Fiz.\  {\bf 79}, 3 (2004)]
  [hep-ph/0311189].
  %%CITATION = HEP-PH/0311189;%%
  %50 citations counted in INSPIRE as of 12 Jun 2014

%\cite{Vogel:2013raa}
\bibitem{Vogel:2013raa} 
  H.~Vogel and J.~Redondo,
  %``Dark Radiation constraints on minicharged particles in models with a hidden photon,''
  JCAP {\bf 1402}, 029 (2014)
  [arXiv:1311.2600 [hep-ph]].
  %%CITATION = ARXIV:1311.2600;%%


\end{thebibliography}
\end{document}